\documentclass[pdflatex,sn-mathphys-num]{sn-jnl} 

\PassOptionsToPackage{colorlinks=true, linkcolor=blue, urlcolor=blue, citecolor=blue}{hyperref}

\UseRawInputEncoding

\usepackage{graphicx}
\usepackage{rotating}
\usepackage{pdfpages}%
\usepackage{multirow}%
\usepackage{amsmath,amssymb,amsfonts}%
\usepackage{amsthm}%
\usepackage{mathrsfs}%
\usepackage[title]{appendix}%
\usepackage{xcolor}%
\usepackage{textcomp}%
\usepackage{manyfoot}%
\usepackage{booktabs}%
\usepackage[ruled,linesnumbered,vlined]{algorithm2e}
\usepackage{algorithmicx}%
\usepackage{algpseudocode}%
\usepackage{listings}%
\usepackage{svg}
\usepackage{tabularx}
\usepackage{booktabs}
\usepackage{array}
\usepackage{ragged2e}
\usepackage{caption}
\usepackage{stfloats}   
\usepackage{makecell}

\usepackage{enumitem}
\usepackage{bm}
\usepackage{relsize}
\usepackage{hyperref}

\theoremstyle{thmstyleone}%
\theoremstyle{thmstyletwo}%

\theoremstyle{thmstylethree}%

\begin{document}
\title[Article Title]{Universal Network Generation Model via Exponential Probabilistic Growth and Vari-linear Preferential Attachment}



\author[1]{\sur{Jinhu Ren}}\email{jinhuren212@mail.ustc.edu.cn}
\author*[1]{\sur{Linyuan L\"u}}\email{linyuan.lv@ustc.edu.cn (Linyuan L\"u)}

\affil*[1]{\orgdiv{School of Cyber Science and Technology}, \orgname{University of Science and Technology of China}, \orgaddress{\street{No. 96 Jinzhai Road}, \city{Hefei}, \postcode{230026}, \state{Anhui}, \country{China}}}


\abstract{
Generated networks are widely used in network-based research as a convenient simulation environment.
Generating universal networks that more accurately reflect real-world patterns is a cornerstone task.
This study proposes a vari-linear network generation model that incorporates two core mechanisms: exponential probabilistic growth and vari-linear preferential attachment.
It concurrently overcomes the limitations of traditional growth in characterizing the low-degree region of the degree distribution and the issues regarding the universality of linear preferential attachment.
Results indicate that our model describes real-world networks more comprehensively and faithfully, and is highly interpretable.
Its performance on diverse empirical datasets is several times better than traditional methods.
Related mechanisms and conclusions are substantiated through ablation experiments and statistical analysis.
Notably, it achieves a unified interpretation of previously isolated classical network characteristics.
This work not only provides a higher-quality universal network generation method, but also bridges the boundaries between traditional concepts, thereby promoting substantive progress in the ``world model'' of networks.
}


\maketitle


\newpage
\section{Introduction}\label{sec1}

Networks are a cornerstone for describing and simulating complex systems in the real world~\cite{RN439}, and are crucial to revealing structural and functional relationships in a wide variety of domains~\cite{RN788}.
Network generation models~\cite{RN779} have long been an indispensable simulation platform in the research of complex networks~\cite{RN195} due to their convenience and low cost.

Classical models such as Watts-Strogatz~\cite{RN279}, Price~\cite{RN690, RN689} and Barab\'{a}si-Albert~\cite{RN105} successfully capture certain typical characteristics of real-world degree distributions.
Among them, the scale-free model proposed a framework consisting of growth and preferential attachment~\cite{RN683}. This framework and its resulting power-law distribution are widely recognized~\cite{RN678, RN227}.
Factors that influence preferential attachment have been further explored~\cite{RN139} such as fitness~\cite{RN685}, homogeneity~\cite{RN684, RN703}, and the Euclidean distance between nodes~\cite{RN701, RN700, RN686}.
In addition, the nonlinear preferential attachment is considered to show potential to generate degree distributions that more closely fit real-world networks~\cite{RN692, RN695, RN707}.
Moreover, the growth of random edge counts is also considered to positively contribute to a better distribution, especially at the low-degree (head) region~\cite{RN777}.
Zadorozhnyi and Yudin~\cite{RN793} proposed a framework allowing new nodes to introduce a random number of edges and to connect via nonlinear preferential attachment rule, and analyzed its theoretical properties.
More recently, learning-based graph generation methods show advantages in mimicking network structures, thereby challenging heuristic methods~\cite{RN789}.

Although existing methods come closer to capturing the characteristics of real-world networks, they still have limitations.
First, the universality of power-law distribution remains contentious~\cite{RN680, RN705}, with an empirical analysis indicating that only a tiny fraction (4\%) of real-world networks exhibit strong power-law~\cite{RN783, RN681}.
Second, optimizing preferential attachment via additional preference factors or hybrid rules increases parameter complexity and often relies on subjective information that is difficult to obtain. 
Third, although there are methods that employ random growth and nonlinear preferential attachment, the probabilistic form of the growth remains unspecified, and its universality in real-world networks lacks extensive validation.
Moreover, learning-based methods exhibit limited interpretability and incur prohibitive computational and memory costs on large networks.
Finally, existing methods often rely on limited empirical data for validation, particularly covering only a narrow range of real-world scenarios.
These issues limit the effectiveness and generality of existing methods in practical applications.

To address these limitations, we propose a vari-linear network generation model (vari-linear model for short) that simultaneously includes two core mechanisms: growth governed by an exponential probability distribution and preferential attachment with variable-linearity.
Through the combination of these two mechanisms, the model can more comprehensively and faithfully capture the degree-distribution patterns of real-world networks, particularly by simultaneously accounting for both the low-degree (head) and high-degree (tail) regions.
Beyond zeroth-order parameters (node and edge counts), the generation process requires only one vari-linear control parameter $r$ for concise control.
As a heuristic, the model is computationally lightweight and readily scales to generate large networks.
Its generalizability is thoroughly validated by utilizing the dozens of real-world networks with wide range of scales and diverse categories as empirical datasets.
Notably, the model provides a unified theoretical explanation for topological characteristics arising from classical models.


\section{Results}\label{sec2}

\subsection{Vari-linear Network Generation Model}\label{sec2-1}

\begin{figure}[t] 
\centering
\includegraphics[width=\textwidth, keepaspectratio]{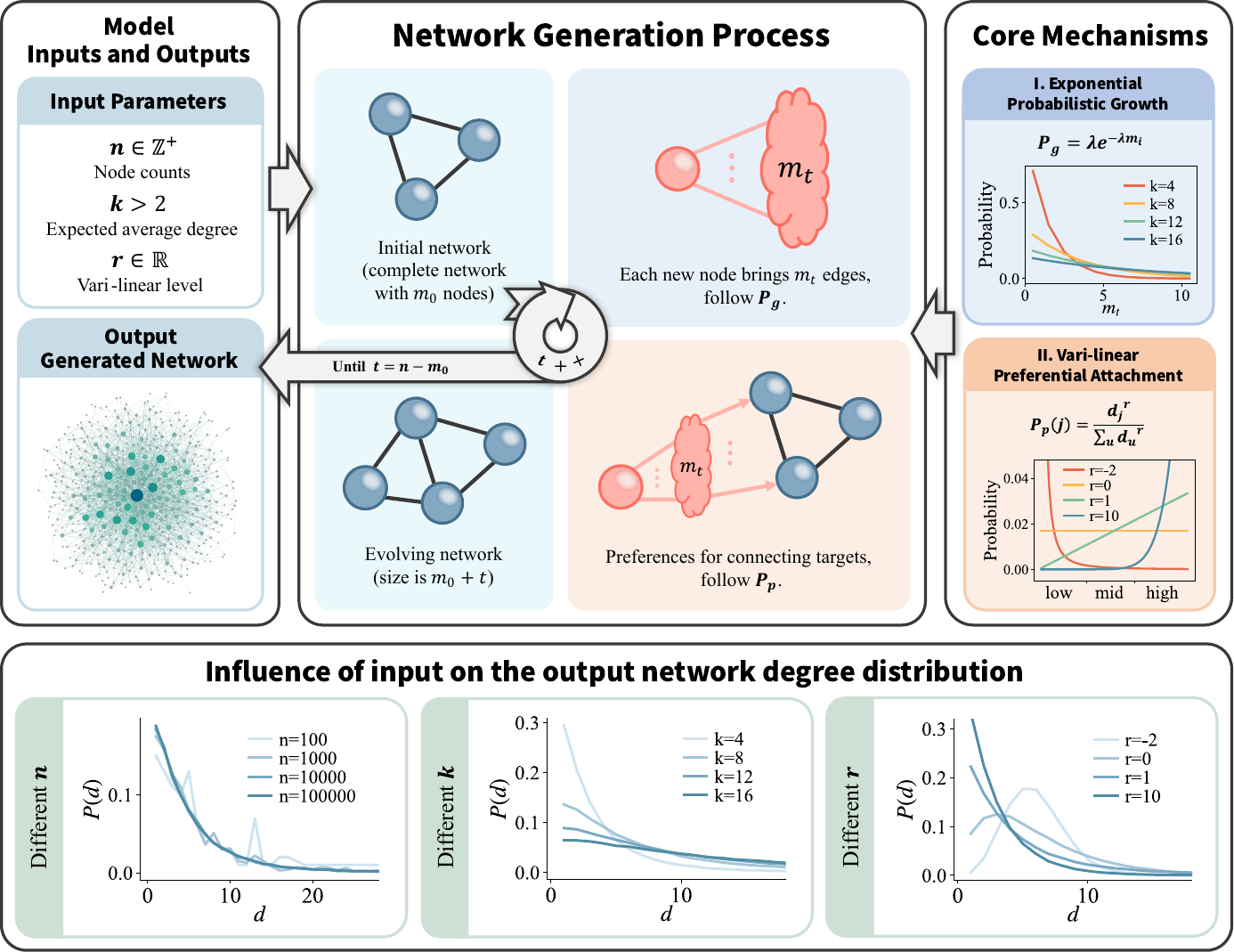}
\caption{
The architecture and mechanisms of vari-linear network generation model.
The model's inputs are the parameters to pre-set and to control the network generation process, and its output is the generated network.
The model generation process iterates by continuously adding new nodes and connecting them to the evolving network.
The model's core mechanisms are exponential probabilistic growth and vari-linear preferential attachment.
The model's parameters exert a controllable influence on the structural characteristics of the output network by regulating the details of the generation process.
}
\label{fig-concept}
\end{figure}

The vari-linear network generation model and its operational principles are shown in Fig.~\ref{fig-concept}. The core steps of the model are as follows.

\paragraph{I. Exponential Probabilistic Growth}

Starting from a seed graph (small complete graph) with $m_0$ nodes, one new node $i$ is added at a time and connected to $m_i$ existing nodes in the graph, where $m_i \geq 1$ and no larger than the number of nodes in the current graph. Repeat until the network size reaches the preset number of nodes $n$ (an input parameter of model). 

Assume the expected average degree of network is $k$ (an input parameter of model), selection probability of the value of $m_i$ follows an exponential distribution:
\begin{equation}
    {P_g} = \lambda {e^{ - \lambda m_i}} ,
\label{eq1}
\end{equation}
to ensure that the average degree of the resulting network approximates $k$, we set $\lambda  =  - \ln (1 - 2/k)$ (as deduced in Eq.~\eqref{eq4}-\eqref{eq11}).
Thus, the complete expression is:
\begin{equation}
    {P_g}(X = {m_i}) 
    = - \ln \left( {1 - \frac{2}{k}} \right) {e^{ \ln \left( {1 - \frac{2}{k}} \right) m_i}}
    = - \ln \left( {1 - \frac{2}{k}} \right) \left( {1 - \frac{2}{k}} \right)^{m_i} .
\label{eq2}
\end{equation}

\paragraph{II. Vari-linear Preferential Attachment}

For each newly added node, the probability that it is connected to an existing node $j$ in the network is:
\begin{equation}
    {P_p}\left( j \right) = \frac{{{d_j}^r}}{{\sum\nolimits_u {{d_u}^r} }},
\label{eq3}
\end{equation}
where $d_j$ denotes the degree of the node $j$, and $r$ is the parameter controlling the level of variable-linearity for preferential attachment (an input parameter of model). 
Note the distinction between $k$ and $d$: $k$ denotes the parameter value input into the model representing the expected average degree, while $d$ denotes the actual node degree value in networks. 

The computational procedure for the model is as described in Supplementary Information (SI) Appendix~\ref{A-code}.

\subsection{Real-world Network Datasets for Validation}\label{sec2-2} 

The experiments in this work use $32$ publicly available real-world network datasets. The size of networks spans three orders of magnitude, and cover a wide range of real-world scenarios, such as social, communication, scholarly co-authorship, academic citation, biology, literature and art. Full details are provided in SI-Appendix~\ref{A-ds}.

\subsection{Empirical Validation of Network Similarity}\label{sec2-3}

\begin{figure}[t] 
\centering
\includegraphics[width=\textwidth, keepaspectratio]{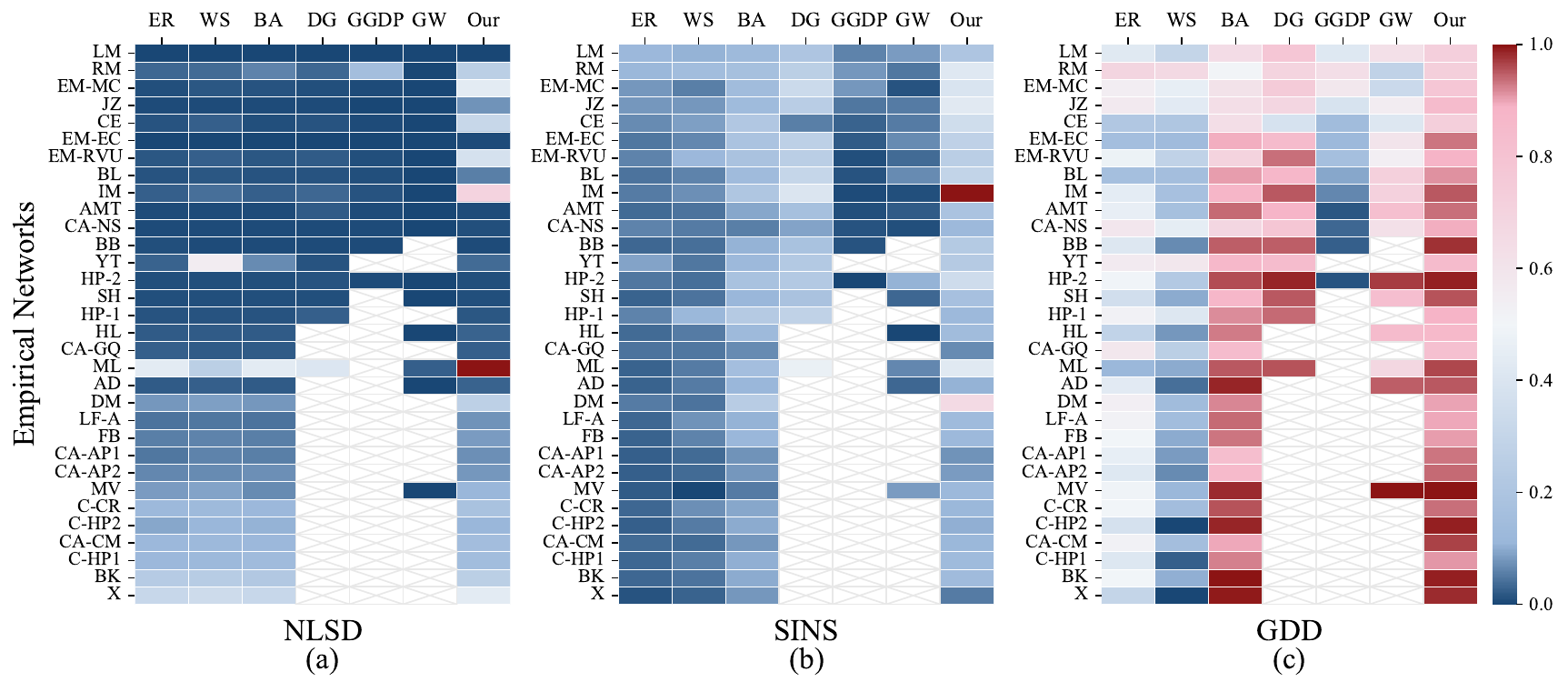}
\caption{
Heatmaps of similarity between model-generated networks and their corresponding real-world networks.   
Subgraphs show the results for three network similarity metrics: (a) Network Laplacian Spectral Descriptor (NLSD), (b) Size-Independent Network Similarity (SINS), and (c) Graphlet Degree Distribution (GDD). 
As NLSD and SINS are distance-like metrics (where lower value indicates higher similarity), take their inverse here.
Gray cells indicate cases where learning-based models are unable to perform computations even under the conditions of our high-performance computing platform.
Detailed numerical results are provided in SI-Appendix~\ref{A-ns}.
}
\label{fig-ns}
\end{figure}

Through three complementary similarity metrics, we provide a comprehensive assessment of each model's ability to reproduce real-world structural patterns, as shown in Fig.~\ref{fig-ns}. 
Across all similarity metrics, the vari-linear model exhibits consistently higher similarity to real-world networks than existing models on the majority of datasets. 

On NLSD metric, it achieved average improvements of $2.35\times$, $1.97\times$ and $2.28\times$ compared to the ER, WS and BA models, respectively. And even larger advantages over learning-based models, including $3.41\times$ on DG, $6.66\times$ on GGDP and a remarkable $193.84\times$ on GW. 
GW's poor performance stems from its reliance on short-horizon, heavily smoothed random-walk trajectory constraints. These constraints fail to control the full Laplacian spectrum and therefore lead to substantial deviations from real networks under the NLSD heat-kernel spectral metric.
On SINS metric, it achieved average improvements of $1.90\times$, $1.77\times$, $1.32\times$ and $1.17\times$ compared to the ER, WS, BA and DG models, respectively. And maintains clear advantages over GGDP ($2.83\times$) and GW ($2.33\times$). 
On GDD metric, it achieved average improvements of $2.16\times$, $6.15\times$ and $1.06\times$ compared to the ER, WS and BA models, respectively. Compared with learning-based models, it achieved average improvements of $1.07\times$ on DG, $1.37\times$ on GW and a remarkable $6.97\times$ on GGDP.
These conclusions are statistically stable, as all 95\% confidence-interval lower bounds exceed $1$, and statistically consistent, with all Wilcoxon tests yielding $p<5\times10^{-2}$ (often far smaller) except for the comparison with DG ($p=5.05\times10^{-2}$).
Related numerical results are provided in SI-Appendix~\ref{A-ns} and SI-Appendix~\ref{A-sa}.

Thus, the results indicate that the vari-linear model more faithfully preserves the structure of real-world networks than existing methods and demonstrates strong generality.

\begin{figure}[t] 
\centering
\includegraphics[width=\textwidth, keepaspectratio]{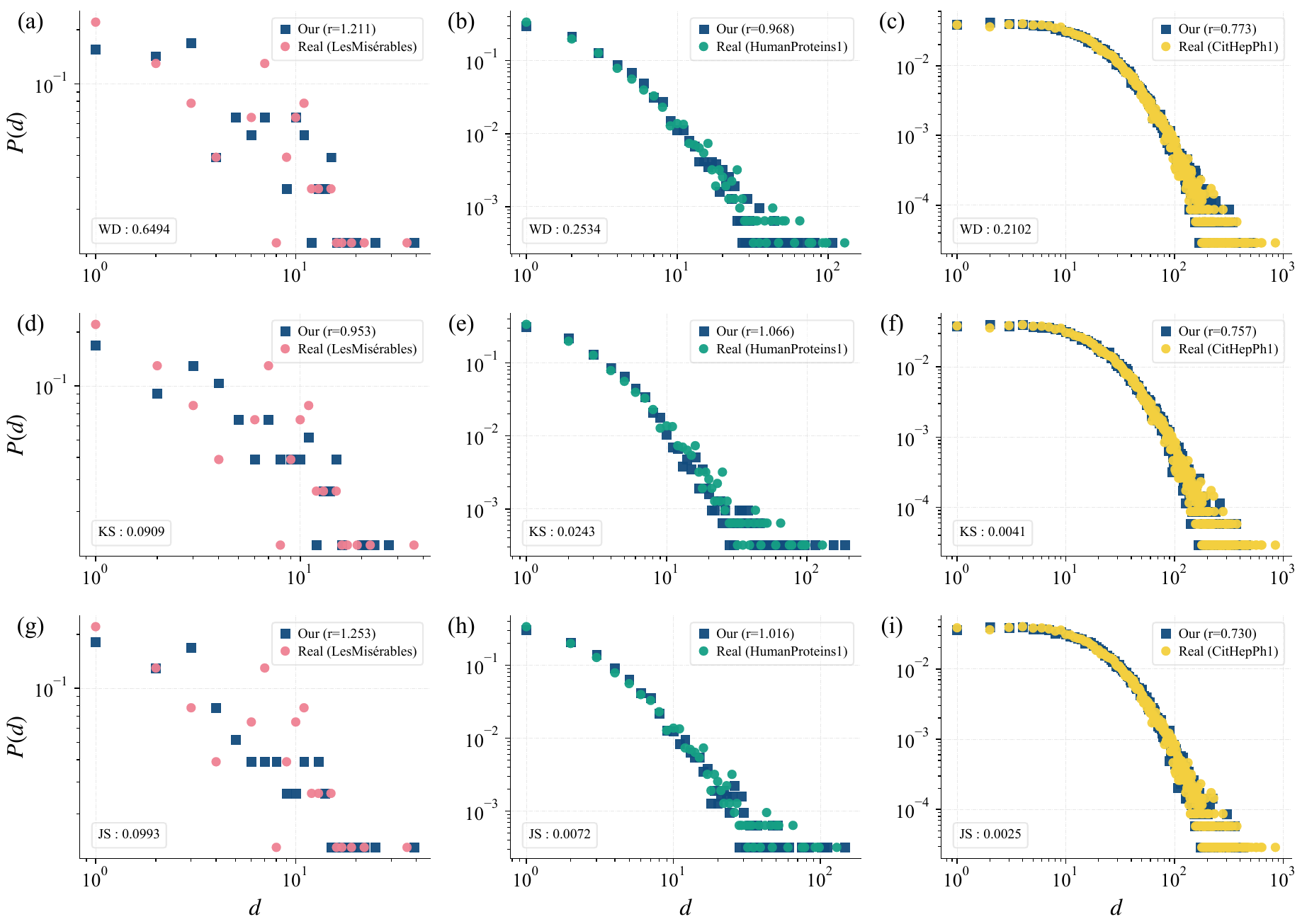}
\caption{
Comparison of the degree distribution probabilities (log-log scale) between vari-linear model-generated networks and real-world networks.
x-axis denotes degree $d$, the number of edges connected to node. y-axis denotes probability $P(d)$, the proportion of nodes with degree $d$.
The three rows of subplots correspond to the optimal results under the three degree distribution divergence metrics: (a-c) Wasserstein Distance (WD), (d-f) Kolmogorov-Smirnov (KS) test, (g-i) Jensen-Shannon (JS) divergence.
The three columns of subplots represent networks of different categories and scales:
pink denotes the $LesMis\acute{e}rables$ network, a small network with $77$ nodes belonging to the $Literature \& Art$;
green denotes the $HumanProteins1$ network, a medium network with $3,133$ nodes belonging to the $Biological$;
yellow denotes the $CitHepPh1$ network, a large network with $34,546$ nodes belonging to the $Citation$.
Results for the model across all $32$ real-world networks are provided in SI-Appendix~\ref{A-rw-dd}.
}
\label{fig-dd}
\end{figure}

\subsection{Empirical Validation of Degree Distribution Divergence}\label{sec2-4}

The distribution of node degrees has long been regarded as a defining characteristic of networks~\cite{RN710}. 
We compared the degree distribution patterns of the networks generated by models with those of real-world networks.
Across various metrics of degree distribution divergence, the vari-linear model consistently achieves high goodness-of-fit on most real-world networks, partial results are shown in Fig.~\ref{fig-dd}.

On WD metric, it achieved average improvements of $6.23\times$, $7.07\times$ and $3.60\times$ compared to the ER, WS and BA models, respectively. Compared with learning-based models, it achieved average improvements of $2.33\times$ on DG, $6.85\times$ on GW and a remarkable $135.17\times$ on GGDP.
On KS metric, it achieved average improvements of $8.67\times$, $10.36\times$ and $8.06\times$ compared to the ER, WS and BA models, respectively. Compared with learning-based models, it achieved average improvements of $2.14\times$ on DG, $11.00\times$ on GGDP and $9.61\times$ on GW. 
On JS metric, it achieved average improvements of $12.58\times$, $17.83\times$ and $11.32\times$ compared to the ER, WS and BA models, respectively. Compared with learning-based models, it achieved average improvements of $1.69\times$ on DG, $12.00\times$ on GGDP and $8.63\times$ on GW.
These conclusions are statistically stable, as all 95\% confidence-interval lower bounds exceed $1$, and statistically consistent, with all Wilcoxon tests yielding $p<5\times10^{-2}$ (often far smaller).
Related numerical results are provided in SI-Appendix~\ref{A-dm-dd} and SI-Appendix~\ref{A-sa}.

\begin{figure}[b]
\centering 
\includegraphics[width=\textwidth, keepaspectratio]{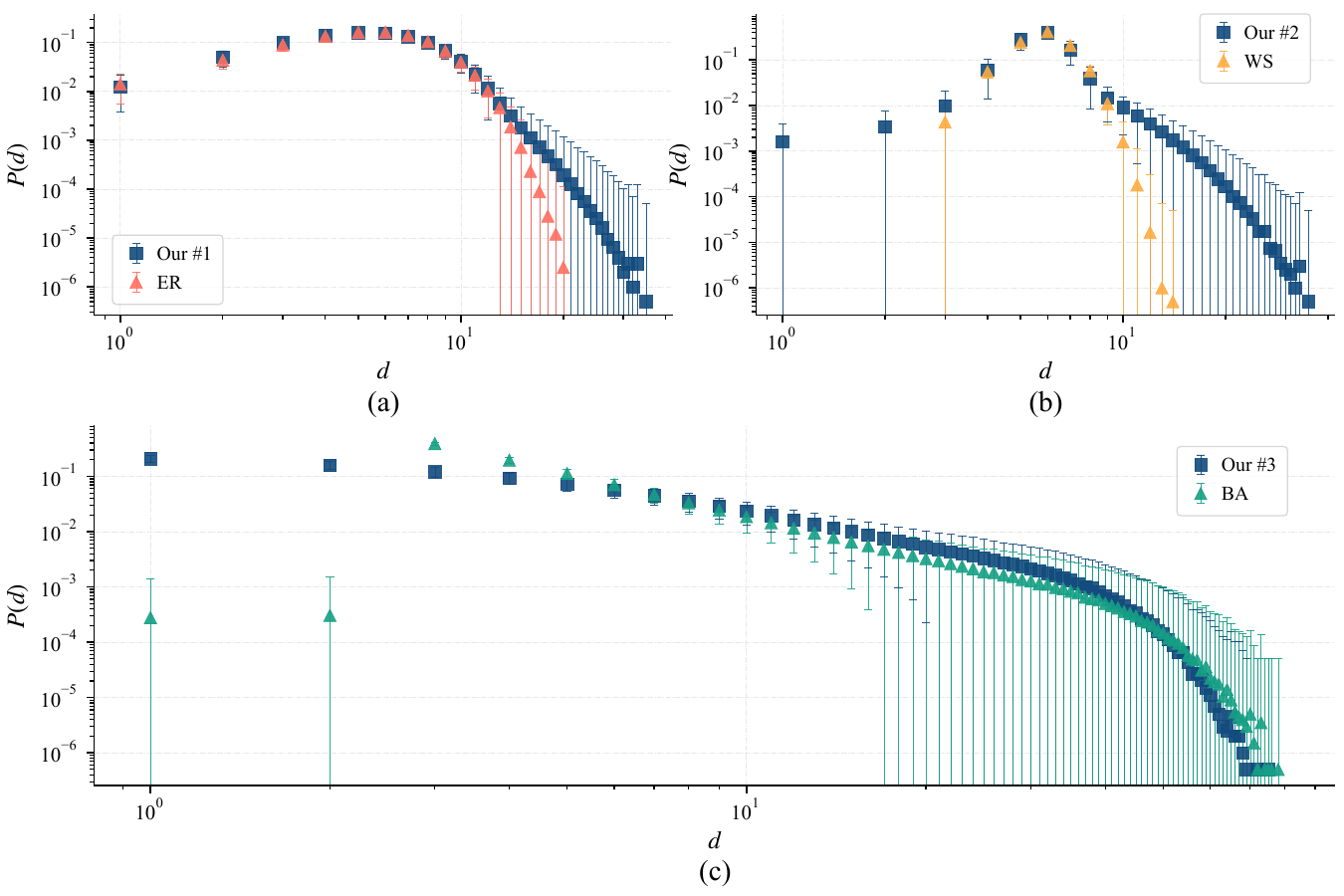}
\caption{
Comparison of vari-linear model with classical distribution of networks (log-log scale).
x-axis denotes degree $d$, the number of edges connected to node. y-axis denotes probability $P(d)$, the proportion of nodes with degree $d$.
(a) Represented by ER network, the non-heterogeneous and short-tailed distribution characteristics resulting from uniform random connections (approximating Poisson or binomial distribution).
(b) Represented by WS network, the nearly normal and uniform distribution characteristics resulting from random reconnection.
(c) Represented by BA networks, the heterogeneous and power-law distribution characteristics resulting from preferential attachment.
All models generate at the scale of $n=200$. The vari-linear networks \#1--\#3 use $k=6.126$ and $r=-1.5,-10,1$, respectively; the ER network uses $p=3\times10^{-2}$; the WS network uses $k=6$ and $p=0.2$; and the BA network uses $m=3$. 
These parameter settings ensure a key baseline that all networks are generated with comparable zero-order properties, meaning the similar scale and average degree.
The experiment is based on $10,000$ Monte Carlo simulations.
}
\label{fig-cn} 
\end{figure}

Thus, the results indicate that the vari-linear model is able to more closely approximate the true degree distribution patterns of real-world networks, thereby offering a more credible theoretical explanation for the emergence of their heterogeneous and heavy-tailed degree structures.

\subsection{Adaptability to Classical Network Characteristics}\label{sec2-5} 

Interpretive rules provided by classical models for the organization of real-world systems are often more significant than those offered by their mechanism frameworks.
To highlight the broader explanatory capacity of the vari-linear model, we conduct comparative experiments with representative classical models of several hallmark network characteristics, as shown in Fig.~\ref{fig-cn}. 
The results show that simply by fine-tuning the parameter $r$ of the vari-linear model to reproduce diverse classical distributional characteristics with minuscule error: the non-heterogeneous and short-tailed distribution arising from random network (Fig.~\ref{fig-cn}a), the nearly normal distribution arising from the small-world network (Fig.~\ref{fig-cn}b), the heterogeneous and power-law distribution arising from scale-free network (Fig.~\ref{fig-cn}c). 
Notably, the uniformity and small-world characteristics of the vari-linear model are primarily evident in the low and medium degree region, whereas the high-degree region always remains influenced by the long-tail effect.
And in terms of network basic properties, vari-linear networks also exhibit marked consistency with their corresponding classical networks.
Specifically, its ability to correspondingly exhibit: the low clustering coefficient arising from random network; the high clustering coefficients at similar levels of average path length arising from small-world network; the short average path length arising from scale-free network.
Related numerical results are provided in SI-Appendix~\ref{A-cn}.

Thus, the results indicate that the vari-linear model can universally generalize the characteristics exhibited by previously isolated classical network models.
And effectively overcomes the limitations of traditional non-power-law models in capturing long-tail phenomenon (as the tail regions of Fig.~\ref{fig-cn}a-b) and the deficiencies of growth of fixed edge count methods in representing low-degree nodes (as the $k < m = 3$ regions of Fig.~\ref{fig-cn}c).

\subsection{Mechanism Contribution Analysis via Ablation}\label{sec2-6}

\begin{figure}[t]
\centering 
\includegraphics[width=\textwidth, keepaspectratio]{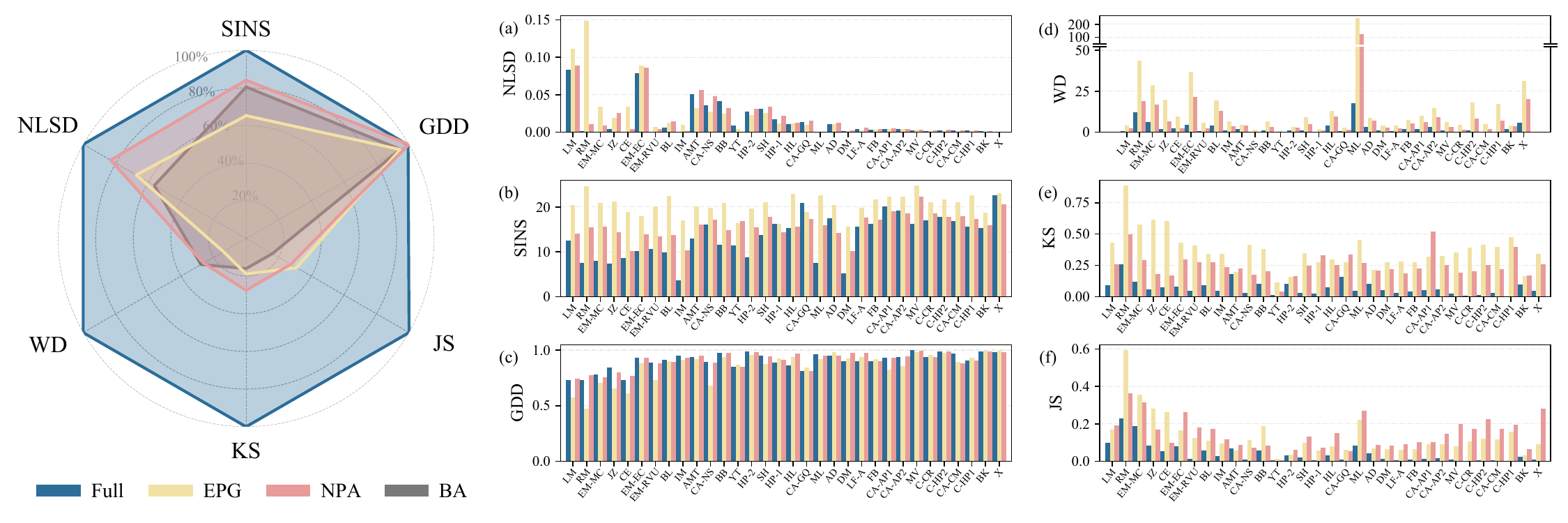}
\caption{
Ablation experiment comparing the full vari-linear model with two ablated variants: Exponential Probabilistic Growth (EPG, mechanism I only) model and Nonlinear Preferential Attachment (NPA, mechanism II only) model, and the baseline BA model. Left: radar plot summarizing mean performance across six metrics (NLSD, SINS, GDD, WD, KS, JS). Right:  bar plots of six metrics results in $32$ real-world networks. For each dataset the generated networks were matched to the empirical networks in size and average degree to ensure comparability. Related numerical results are provided in SI-Appendix~\ref{A-ae}.
}
\label{fig-ae} 
\end{figure}

To assess the necessity of the two core mechanisms in the vari-linear model, we constructed two ablated variants on the BA backbone: exponential probabilistic growth model (EPG, retaining only Mechanism I) and nonlinear preferential attachment model (NPA, retaining only Mechanism II). All models are fitted to 32 empirical networks, and the performance is evaluated using network similarity metrics and degree distribution divergence metrics. 
As shown in Fig.~\ref{fig-ae}, across nearly all datasets and metrics, the full model consistently outperforms both ablated variants. The ablated models exhibit complementary but incomplete strengths: EPG and NPA occasionally approach the full model on individual metrics or datasets, yet neither reproduces the full model's balanced performance across both structural similarity and degree distribution fit. Complete comparisons (across all empirical datasets) yield systematic positive differences in favor of the full model.

The ablation results demonstrate that adopting only a single mechanism leads to systematic degradation in performance. In contrast, the full vari-linear model consistently achieves superior results by combining both mechanisms. This evidence confirms that neither mechanism alone is sufficient, the synergy of two mechanisms in our model is indispensable for faithfully describing diverse real-world networks. 

\subsection{Parameter Sensitivity and Structural Response}\label{sec2-7} 

We perform sensitivity analysis to assess the influence of input parameters in the vari-linear model on the structural characteristics of the generated network.
The definitions of the input parameters are summarized in Table~\ref{tab-se}. We conducted systematic simulations under different parameter input schemes and analyzed both the fundamental network properties and the degree distribution patterns of the resulting networks.

\begin{figure}[t]
\centering 
\includegraphics[width=\textwidth, keepaspectratio]{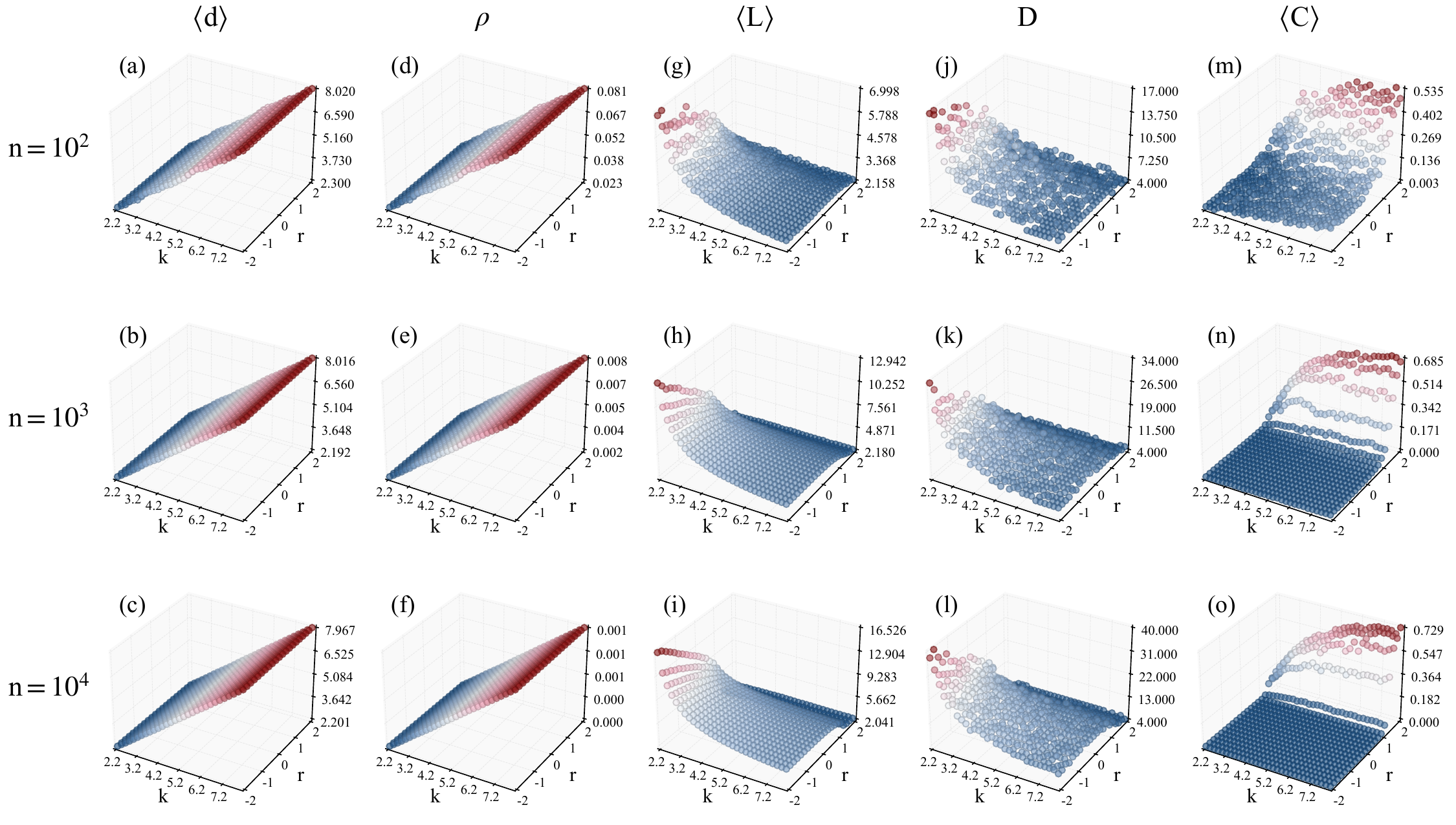}
\caption{
Different parameter configurations of the vari-linear model and the basic properties of the corresponding output network. Each row corresponds to network scale from small to large ($n=100$, $n=1000$, $n=10000$), while each column represents a specific structural propertie of the resulting networks: average degree $\langle d \rangle$, density $\rho$, average path length $\langle L \rangle$, diameter $D$, and average clustering coefficient $\langle C \rangle$. Within each 3D subplot, the $x$-axis denotes the model parameter $k$, the $y$-axis denotes the model parameter $r$, and the $z$-axis denotes the corresponding property value.
}
\label{fig-pa-3d} 
\end{figure}

Five canonical properties are examined, results reveal clear and consistent trends, as shown in Fig.~\ref{fig-pa-3d}. The average degree and density increase linearly with $k$ and remain essentially independent of $r$. In contrast, both average path length and diameter decrease with increasing $k$ and $r$, reflecting denser connectivity and enhanced global reachability. The clustering coefficient shows a positive correlation with $k$ and $r$, but this effect becomes pronounced only under super-linear preferential attachment ($r>1$), where the exponential influence of $r$ amplifies local aggregation. Thus, these observations indicate that $k$ exerts a significant influence on global structural properties defined by the edge count, while $r$ directly influences the global aggregation level of connections. These patterns are robust across different network sizes, with the trends becoming more pronounced as scale increases.

\begin{figure}[t]
\centering 
\includegraphics[width=\textwidth, keepaspectratio]{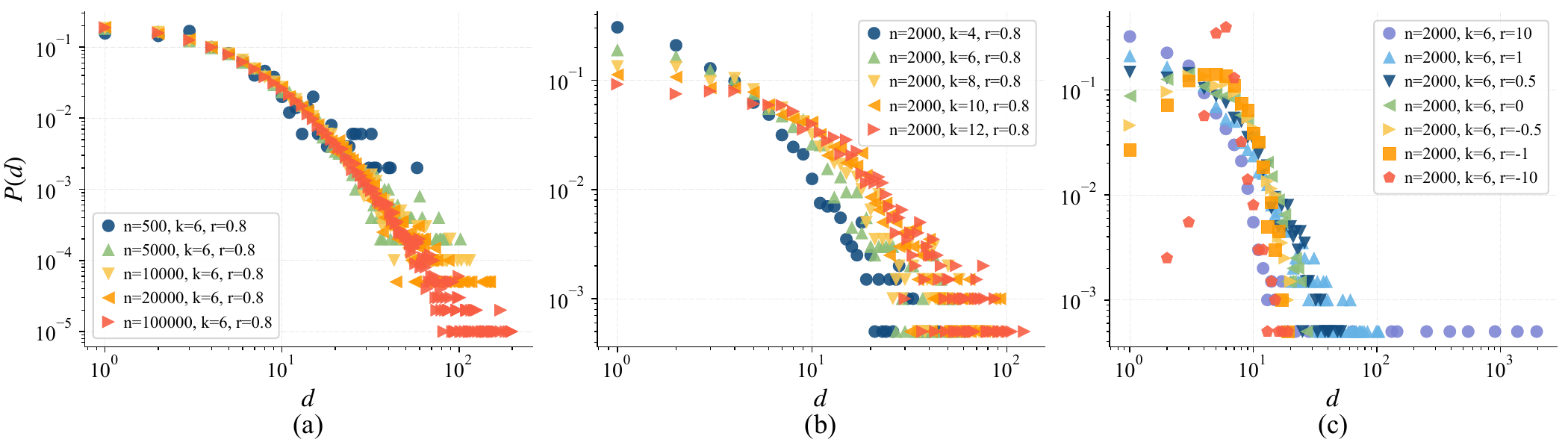}
\caption{Degree distribution probability (log-log scale) of the vari-linear model under typical parameter values.
x-axis denotes degree $d$, the number of edges connected to node. y-axis denotes probability $P(d)$, the proportion of nodes with degree $d$.
(a) Resulting network degree distributions for different number of nodes parameter $n$. (b) Resulting network degree distributions for different average degree parameters $k$. (c) Resulting network degree distributions for different vari-linear parameters $r$.
}
\label{fig-pa-dd} 
\end{figure}

We further analyzed the degree distributions under typical parameter settings (Fig.~\ref{fig-pa-dd}). 
The stability of structural characteristics under large variations in network size (Fig.~\ref{fig-pa-dd}a) demonstrates that the vari-linear model possesses strong scale invariance, structural robustness, and stable scalability. 

Variation in the expected average degree $k$ directly governs the number of new edges introduced by incoming nodes, thereby shaping the distribution of the region around $k$ degrees (primarily in low to medium region). 
With increasing $k$ (Fig.~\ref{fig-pa-dd}b), the stochastic growth mechanism ensures that degrees smaller than $k$ do not vanish, instead gradually decline at the defined probability. 
Meanwhile, when preferential attachment is positively correlated with degree ($r > 0$), high-degree nodes progressively attract more connections during the evolutionary process, due to more new edges are added.
This reinforces the accumulation effect of the edges, thereby raising the tail of the distribution.

The parameter $r$ exerts decisive influence on distributional form. 
Super-linear attachment ($r \gg 1$) produces networks dominated by the vast majority of low-degree nodes and a few hyper-high-degree nodes, resembling winner-take-all phenomenon. 
Linear attachment ($r = 1$) produces distribution similar to power-law, yet primarily evident in medium-to-high region.
Sub-linear attachment ($0 < r < 1$) weakens the heterogeneity and clustering of the network, thereby regulating long-tail regions.
Smaller values ($r \leq 0$) further equalize connectivity, and for $r \ll 0$ the distribution concentrates tightly around the average degree, with hubs essentially absent. 
In summary, larger $r$ values drive centralization and hub dominance, while smaller or negative $r$ values promote decentralization and fairness. 
Note that due to the probabilistic growth, the overall distribution of the network is not strictly determined by the preferential attachment.

Thus, the results indicate that the fundamental properties and degree distribution patterns of networks generated by the vari-linear model respond to parameter variations in a systematic and interpretable manner. The joint influence of $k$ and $r$ provides a coherent mechanism for shaping structural outcomes, and these capabilities are reliably reproduced across diverse network scales.

\section{Discussion}\label{sec3}

This work proposes a vari-linear network generation model.
Its performance is assessed by comparing it with several classical network models and recent learning-based generation models, using multidimensional metrics that quantify network similarity and degree distribution divergences.
To insure generality, the empirical datasets covers $32$ publicly available real-world networks spanning three orders of scale and six diverse actual categories.
Results show that the model produces networks whose structural organization and degree distribution patterns more closely resemble realistic networks.
And, it shows several-fold advantages across majority of datasets and metrics, highlighting its strong structural fidelity and robustness in modeling diverse real-world networks.
And, the contributions of exponential probabilistic growth to the low-degree (head) region and of vari-linear preferential attachment to the high-degree (tail) region are effectively confirmed by the model.
Notably, the model naturally accommodates characteristics previously treated as separate in classical networks, thereby providing a new paradigm that bridges the boundaries between traditional concepts.
The necessity and reliability of the mechanism design of the model are fully validated through ablation experiments and parameter analysis, and its interpretability marked surpasses that of learning-based methods.

The vari-linear model thus provides a unified theory for understanding the diverse characteristics of real-world networks. This also provides a higher-quality foundational simulation environment for downstream research that relies on network (graph) structures. 
In future, examining the factors that influence the level of vari-linear preferences and uncovering their sociological drivers is meaningful.

\section{Methods}\label{sec4}

This section describes the modeling details, statistical analysis methods, and the baseline models, fitting methods, and multidimensional quantitative metrics used in this work. The explanation of all symbols is provided in Table~\ref{tab-se}.

\begin{table}[h]
\centering
\captionsetup{justification=raggedright,singlelinecheck=false} 
\caption{Symbols Explanation.}
\label{tab-se}
\renewcommand{\arraystretch}{1.2} 
\begin{tabular}{>{\centering\arraybackslash}m{1.2cm} l m{7.6cm}}   
\toprule    
\textbf{Symbol} & \textbf{Role} & \textbf{Description} \\
\midrule    
$n$ & Input parameter & The expected node count of the model's output network \\
$k$ & Input parameter & The expected average degree of the model's output network \\
$r$ & Input parameter & The level of variable-linearity for in the model \\
$m_0$ & Initialization parameter & The size of initial seed graph (Default is $3$) \\
$m_i$ & Process variable & The number of edges added by new node $i$ during growth \\
$d_j$ & Process variable & The current degree value of node $j$ \\
$P(d)$ & Basic property &  The probability of a node's degree is $d$ in network \\
$\langle d \rangle$ & Basic property & The average of all nodes' degrees in network \\
$\rho$ & Basic property & The density of network \\
$\langle L \rangle$ & Basic property & The average path length of network \\
$D$ & Basic property & The diameter of network \\
$\langle C \rangle$ & Basic property & The average clustering coefficient of network \\
\bottomrule     
\end{tabular} 
\end{table}

\subsection{Theoretical Analysis}\label{sec4-2}

\subsubsection{Average Degree}\label{sec4-2-1}

The average degree of the vari-linear model is closely related to the settings in step I of the modeling process. Since in practice $m_i \in \{ 1, 2, \dots, N-m_0 \}$, we employ the discrete exponential distribution for more rigorous reasoning. The number of new edges added to node $i$ can be expressed by a probability mass function as:
\begin{equation}
    {P_g}(X = {m_i}) = \frac{{{e^{ - \lambda {m_i}}}\left( {1 - {e^{ - \lambda }}} \right)}}{{{e^{ - \lambda }}\left( {1 - {e^{ - \lambda \left( {N - {m_0}} \right)}}} \right)}},
\label{eq4}
\end{equation}
then the expected value of the number of new edges brought by the new node is:
\begin{equation}
    E[{m_i}] = \sum\limits_{m = 1}^{N - {m_0}} m P_g(m) ,
\label{eq5}
\end{equation}
substituting Eq~\eqref{eq4} yields:
\begin{equation}
    E[{m_i}] = \frac{{1 - {e^{ - \lambda }}}}{{{e^{ - \lambda }}(1 - {e^{ - \lambda \left( {N - {m_0}} \right)}})}}\sum\limits_{m = 1}^{N - {m_0}} m {e^{ - \lambda m}},
\label{eq6}
\end{equation}
using geometric series summation can be derived and simplified to obtain:
\begin{equation}
    E[{m_i}] = \frac{{1 - (N - {m_0} + 1){e^{ - \lambda \left( {N - {m_0}} \right)}}}}{{(1 - {e^{ - \lambda }})(1 - {e^{ - \lambda \left( {N - {m_0}} \right)}})}} + \frac{{\left( {N - {m_0}} \right){e^{ - \lambda (N - {m_0} + 1)}}}}{{(1 - {e^{ - \lambda }})(1 - {e^{ - \lambda \left( {N - {m_0}} \right)}})}}.
\label{eq7}
\end{equation}
For the global average degree $\langle d \rangle$ of the network:
\begin{equation}
    \langle d \rangle  = \frac{{2\left( {N - {m_0}} \right)E\left[ {{m_i}} \right]}}{N} ,
\label{eq8}
\end{equation}
when $N \gg m_0$, then:
\begin{equation}
    \langle d \rangle \approx 2E[{m_i}] .
\label{eq9}
\end{equation}
To ensure that the expected average degree $k$ (input parameter) matches the global average degree in the large-network limit ($N \gg m_0$), we require:
\begin{equation}
k = \langle d \rangle \approx 2E[{m_i}]  =  2\frac{{1 - \left( {N + 1} \right){e^{ - \lambda N}} + N{e^{ - \lambda \left( {N + 1} \right)}}}}{{\left( {1 - {e^{ - \lambda }}} \right)\left( {1 - {e^{ - \lambda N}}} \right)}} .
\label{eq10}
\end{equation}
Thus, to make the average degree of the resulting network close to the expected average degree $k$, should be set:
\begin{equation}
    \lambda  =  - \ln \left( {1 - \frac{2}{k}} \right) .
\label{eq11}
\end{equation}

\subsubsection{Degree Distribution}\label{sec4-2-2}

Since the modeling mechanism of vari-linear model, changes in node degrees in the network are caused by three scenarios. For nodes with degree $d$, there exist (i) Inflow: New edges connect to nodes with degree $d - 1$, increasing their degree to $d$; (ii) Outflow: New edges connect to nodes with degree $d$, increasing their degree to $d + 1$; (iii) Growth: The degree of the newly added node equals $d$.

Let the number of nodes with degree $d$ at step $t$ be $N_{d} (t)$ , then:
\begin{equation}
{N_d}(t + 1) - {N_d}(t) = {N_{d - 1}}(t) {m_t} \frac{{{{(d - 1)}^r}}}{{\sum\nolimits_u {{d_u}^r} }} - {N_d}(t) {m_t} \frac{{{d^r}}}{{\sum\nolimits_u {{d_u}^r} }} + {P_g}(X = d),
\label{eq12}
\end{equation}
where $m_t$ denotes the number of new edges added at step $t$, and $P_g$ is defined in Eq.~\eqref{eq2}-\eqref{eq3}. The first item indicates an inflow, the second item indicates a outflow, and the third item indicates growth.
The sum of vari-linear preferential attachment weights at step $t$ is:
\begin{equation}
\sum\nolimits_u {{d_u}^r}  = \sum\limits_{d = 1}^{t - 1} {{d^r}{N_d}(t)} ,
\label{eq13}
\end{equation}
since:
\begin{equation}
N_d(t) = t{p_d}(t) ,
\label{eq14}
\end{equation}
then:
\begin{equation}
\sum\nolimits_u {{d_u}^r}  = t\sum\limits_{d = 1}^{t - 1} {{d^r}{p_d}(t)} .
\label{eq15}
\end{equation}
Substituting Eq.~\eqref{eq14}-\eqref{eq15} into Eq.~\eqref{eq12} yields the evolution equation for the degree distribution:
\begin{equation}
(t + 1){p_d}(t + 1) - t{p_d}(t) = \frac{{{m_t}}}{{  \sum\limits_{d = 1}^{t - 1} {d^r}{p_d}(t) }} [ {(d - 1)}^r{p_{d - 1}}(t) - {d^r}{p_d}(t) ] + {P_g}(d) .
\label{eq16}
\end{equation}
Under the Large-$t$ approximation condition, it is reasonable to assume that the empirical degree distribution follows:
\begin{equation}
p_d = {p_d}({t \to \infty }) ,
\label{eq17}
\end{equation}
and that the number of new edges added at each step can be replaced by its expectation:
\begin{equation}
m_t \approx E[{m_t}] = \frac{k}{2} .
\label{eq18}
\end{equation}
and the normalization term in Eq.~\eqref{eq16} satisfies:
\begin{equation}
\lim_{t\to\infty}\sum_{d=1}^{t-1} d^r p_d(t)
= \sum_{d\ge 1} d^r p_d .
\label{eq19}
\end{equation}
In the stationary limit condition, it can be considered that:
\begin{equation}
(t+1)p_d(t+1) - t p_d(t) \approx 0 .
\label{eq20}
\end{equation}
The $r$-weighted moment of the stationary distributionn can be expressed as:
\begin{equation}
S_r = \sum_{d\ge 1} d^r p_d.
\label{eq21}
\end{equation}
It represents the total preferential attachment weight of the network in the steady state ($t\to\infty$). Note, $S_r$ is a single scalar that depends on the entire distribution $p_d$, but it does not depend on the degree $d$ itself. Thus, for the moment, it is sufficient to treat $S_r$ as a finite normalization constant characterizing the global strength of vari-linear preferential attachment.

Combining Eq.~\eqref{eq16}-\eqref{eq21} and taking the large-$t$ stationary limit, the explicit dependence on $t$ in Eq.~\eqref{eq16} disappears and we obtain the time-independent first-order linear non-homogeneous recurrence:
\begin{equation}
p_d = \frac{(d-1)^r}{d^r} p_{d-1}  + \frac{2S_r}{k} \cdot \frac{1}{d^r} P_g(d),
\label{eq22}
\end{equation}
where Eq.~\eqref{eq22} is already written in the stationary limit: all quantities $p_d$, $S_r$, $k$, and $P_g(d)$ are time-independent, so the second term does not carry $t$.
The coefficient of the recursive formula is:
\begin{equation}
a_d = \frac{(d-1)^r}{d^r},
\qquad
b_d = \frac{2S_r}{k} \cdot \frac{1}{d^r} P_g(d).
\label{eq23}
\end{equation}
The homogeneous recursive form is:
\begin{equation}
p_d^{(h)}=\frac{(d-1)^r}{d^r}p_{d-1}^{(h)} ,
\label{eq24}
\end{equation}
iterating from $2$ to $d$ gives:
\begin{equation}
p_d^{(h)} = p_1^{(h)} \prod_{u=2}^d a_u 
= p_1^{(h)}\prod_{u=2}^d\frac{(u-1)^r}{u^r}
= p_1^{(h)}\left( \prod_{u=2}^d\frac{u-1}{u} \right)^r ,
\label{eq25}
\end{equation}
where the product in Eq.~\eqref{eq25} simplifies due to its telescoping structure:
\begin{equation}
\prod_{u=2}^d\frac{u-1}{u}=\frac{1}{d} , 
\label{eq26}
\end{equation}
substituting Eq.~\eqref{eq26} into Eq.~\eqref{eq25} yields:
\begin{equation}
p_d^{(h)} = p_1^{(h)} d^{-r} .
\label{eq27}
\end{equation}
Since Eq.~\eqref{eq22} is a first-order linear non-homogeneous recurrence, its general solution follows the standard form:
\begin{equation}
p_d = \left( \prod_{u=2}^d a_u \right)p_1 + \sum_{w=2}^d \left( b_w \prod_{u=w+1}^d a_u \right).
\label{eq28}
\end{equation}
Using Eq.~\eqref{eq25} and Eq.~\eqref{eq27} yields:
\begin{equation}
\prod_{u=2}^d a_u = d^{-r} , 
\label{eq29}
\end{equation}
by the same telescoping argument from Eq.~\eqref{eq26}, for any $w<d$:
\begin{equation}
\prod_{u=w+1}^d a_u = \left(\prod_{u=w+1}^d \frac{u-1}{u} \right)^r = \left( \frac{w}{d} \right)^r . 
\label{eq30}
\end{equation}
Substituting Eq.~\eqref{eq23}, Eq.~\eqref{eq29} and Eq.~\eqref{eq30} into Eq.~\eqref{eq28} yields:
\begin{equation}
\begin{aligned}
p_d
&= d^{-r} p_1 + \sum_{w=2}^d \left[ \left(\frac{2 S_r}{k}\cdot\frac{1}{w^r}P_g(w)\right)\left(\frac{w}{d}\right)^r \right] \\
&= d^{-r} p_1 + \frac{2 S_r}{k}\cdot\frac{1}{d^r}\sum_{w=2}^d P_g(w).
\end{aligned}
\label{eq31}
\end{equation}
where the first term reflects the homogeneous decay inherited from the initial condition, while the second term represents the cumulative contribution of all non-homogeneous injections $P_g(w)$ up to degree $d$, each transported to $d$ through the factor $(w/d)^r$ and scaled by the global normalization $2S_r/k$.
Using the exponential growth kernel $P_g$ defined in Eq.~\eqref{eq2}, the
cumulative injection term in Eq.~\eqref{eq31} is expressed as:
\begin{equation}
\sum_{w=2}^d P_g(w)
= \sum_{w=2}^d
  \Bigl[
    - \ln\!\left(1 - \frac{2}{k}\right)
    \left(1 - \frac{2}{k}\right)^{w}
  \Bigr] ,
\label{eq32}
\end{equation}
applying geometric-series summation yields:
\begin{equation}
\begin{aligned}
\sum_{w=2}^d P_g(w)
&= - \ln\!\left(1 - \frac{2}{k}\right)
\left(1 - \frac{2}{k}\right)^{2}
\frac{1 - \left(1 - \frac{2}{k}\right)^{d-1}}{1 - \left(1 - \frac{2}{k}\right)}  \\
&= - \frac{k}{2}
\ln\!\left(1 - \frac{2}{k}\right)
\left(1 - \frac{2}{k}\right)^{2}
\left[1 - \left(1 - \frac{2}{k} \right)^{d-1} \right].
\end{aligned}
\label{eq33}
\end{equation}
Substituting into Eq.~\eqref{eq31} yields the explicit stationary degree distribution:
\begin{equation}
\begin{aligned}
p_d
&= d^{-r} p_1
  + \frac{2 S_r}{k}\, d^{-r}
    \left[
      - \frac{k}{2}
      \ln\!\left(1 - \frac{2}{k}\right)
      \left(1 - \frac{2}{k}\right)^{2}
      \left(1 - \left(1 - \frac{2}{k} \right)^{d-1} \right)
    \right] \\
&= d^{-r}
   \left[
     p_1
     - S_r
       \ln\!\left(1 - \frac{2}{k}\right)
       \left(1 - \frac{2}{k}\right)^{2}
       \left(1 - \left(1 - \frac{2}{k} \right)^{d-1} \right)
   \right].
\end{aligned}
\label{eq34}
\end{equation}

Here, we do not aim to derive an exact closed-form solution for the stationary degree distribution, as this is theoretically difficult to achieve even for classical methods. Although the constants $p_1$ and $S_r$ are ultimately fixed by global normalization, Eq.~\eqref{eq34} nevertheless provides a structurally informative characterization that elucidates how the parameters $r$ and $k$ jointly influence the shape of the stationary degree distribution.

\textbf{Main effects of $k$: control over low-medium (head region) degree distribution.}
Larger $k$ results in slower attenuation of the effects with degree, so its influence persists across a wider range of low and medium degrees. This produces a broader head region and delays the onset of the tail region.
Smaller $k$ results in faster attenuation of the effects with degree. This produces a shorter head region before transitioning into a longer tail region.
Thus, $k$ mainly determines the distribution state in the low to medium degree region, such as the central position and coverage area of the head. Meaning, this confirms that the parameter $k$ determines the intensity and decay rate of the effect produced by the step of growth.

\textbf{Main effects of $r$: control over medium-high (tail region) degree distribution.}
Changing $r$ will simultaneously alter the local decay factor $d^{-r}$ and the global normalization term $S_r$, which is self-consistently determined by the entire stationary distribution. In the self-consistent solution of Eq.~\eqref{eq34}, this feedback can reverse the naive expectation based solely on $d^{-r}$.
Based on this, larger $r$ results in a slower decline of the distribution, forming a heavier-tailed distribution (with more hub nodes).
And, smaller $r$ results in a faster decline in the distribution, and suppresses the distribution region of high degree, producing a more homogeneous network. 
Thus, $r$ mainly determines the distribution state in the medium-high degree region, such as the length of tail. Meaning, this confirms that the parameter $r$ determines the preferential intensity produced by the step of attachment. 


\subsection{Statistical Assessment of Result Disparities}\label{sec4-3}

To ensure that the performance comparison between the proposed vari-linear model and baseline generative models is both rigorous and robust, we employ a three-part statistical evaluation protocol. This protocol quantifies not only the numerical disparities in performance results but also their stability across networks and their statistical significance.

\textbf{Geometric-mean improvement factor}:
For each metric and baseline model, we summarize the per-network performance ratios using the geometric mean. This measure captures the typical multiplicative improvement of RL over each baseline and is well suited for ratio-based comparisons, which are often skewed and multiplicative in nature. A geometric-mean ratio greater than~1 indicates that the model achieves consistently lower error (or higher score for ``higher-is-better'' metrics) across networks.

\textbf{Bootstrap confidence interval}:
To evaluate the stability of the observed improvements, we compute a non-parametric 95\% bootstrap confidence interval for the geometric-mean ratio. This interval reflects the variability of the improvement across the empirical networks without assuming any specific distributional form. A confidence interval whose lower bound exceeds~1 provides strong evidence that the model's advantage is robust and not driven by a small subset of networks.

\textbf{Wilcoxon signed-rank test}:
To assess whether the performance differences are statistically significant on a per-network basis, we apply the Wilcoxon signed-rank test to paired results from our model and each baseline model. This non-parametric test is appropriate for heterogeneous real-world networks and does not rely on normality assumptions. A small $p$-value indicates that the model's superiority is systematic rather than attributable to random fluctuations.

Together, these three components provide a comprehensive evaluation of model reliability: the geometric-mean ratio quantifies the magnitude of improvement, the bootstrap interval assesses its stability, and the Wilcoxon test establishes statistical significance. All results reported in Section~\ref{sec2-3}-\ref{sec2-4} follow this evaluation protocol.

\subsection{Baseline Network Models for Comparison}\label{sec4-4}

This work uses three classical network models, three cutting-edge graph generation methods, and two network models each featuring relevant core mechanisms as baselines for comparison with the vari-linear model.

\subsubsection{Classical Network Models}\label{sec4-4-1}

Three widely recognized classical network models were employed for comparison in this work. 

\textbf{Erd\"{o}s-R\'{e}nyi (ER) Network Model}: The Erd\"{o}s-R\'{e}nyi random network model~\cite{RN226} assumes a probability $p_{ER}$ that any two distinct nodes among $n$ nodes are connected by an edge. Thus, when the number of nodes is fixed, adjusting $p_{ER}$ provides a direct means of controlling the total number of edges in the ER network.

\textbf{Watts-Strogatz (WS) Network Model}: The Watts-Strogatz (WS) small-world network model~\cite{RN279} assumes that each node among $n$ nodes connects to $k$ ($k>2$) nearest neighbors, and edge reconnection occurs with probability $p_{WS}$. Due to the fact that the reconnection step in the model does not alter the total number of edges. Thus, when the number of nodes is fixed, adjusting $k$ provides a direct means of controlling the total number of edges in the WS network.

\textbf{Barab\'{a}si-Albert (BA) Network Model}: The Barab\'{a}si-Albert (BA) scale-free network model~\cite{RN105} generates a network through an iterative growth-attachment process. Starting from a seed graph (empty graph) of $m_{0}$ nodes, the model performs $n - m_{0}$ growth steps, and at each step a new node is introduced and connected to $m$ existing nodes according to the preferential attachment rule. Since each iteration contributes exactly $m$ new edges, the total number of edges in the resulting network is approximately $mn$. Thus, when the number of nodes is fixed, adjusting $m$ provides a direct means of controlling the total number of edges in the BA network.

\subsubsection{Recent Learning-based Graph Generation Models}\label{sec4-4-2}

Recent graph generation methods increasingly rely on learning-based architectures, three cutting-edge graph generation methods were employed for comparison in this work. Related experiments are conducted in a high-performance computing environment equipped with NVIDIA RTX Pro 6000 Blackwell Server Edition GPUs.

\textbf{DiGress (DG)}: 
The DiGress~\cite{RN785} is a discrete denoising diffusion model that applies independent Markovian transitions to categorical node and edge types through time-dependent transition matrices. A graph transformer network is trained to invert this process by predicting clean node and edge categories from noisy graphs, optimized via a cross-entropy loss decomposed over all nodes and edges. The model operates entirely in the discrete domain, preserving graph sparsity throughout diffusion. This discrete formulation improves sample quality and enables better performance on both molecular and non-molecular graph generation tasks.

\textbf{GraphGDP (GGDP)}:
The GraphGDP~\cite{RN786} is a continuous-time generative diffusion framework that perturbs real-valued adjacency matrices via a variance-preserving SDE and generates graphs by solving the corresponding reverse-time SDE. Its forward process admits a closed-form Gaussian perturbation kernel, and its position-enhanced graph score network (PGSN) integrates continuous perturbations with structural and positional features extracted from quantized intermediate graphs for permutation-equivariant score estimation. The framework jointly leverages continuous SDE dynamics and discrete graph structure. It achieves competitive generation quality with substantially fewer function evaluations than autoregressive models.


\textbf{GraphWeave (GW)}:
The GraphWeave~\cite{RN787} generates graphs by first modeling structural patterns through smoothed Random Walk Trajectories (RWTs), and then reconstructing a graph that best fits these trajectories via a global optimization over all edges. A permutation-invariant transformer is trained to reverse RWT steps, enabling the generation of realistic trajectories from analytically derived ending vectors based on random-walk convergence. This separation of pattern generation and graph construction allows the method to capture structural properties reflected in random-walk behavior. It demonstrates significant advantages in metrics involving large-scale structural patterns.


\subsubsection{Models for Ablation Experiment}\label{sec4-4-3}

Two models for comparison are established, respectively equipped with two mechanisms from vari-linear model. The design drew inspiration from ablation experiments, achieving an intuitive comparison of the effectiveness of the two core mechanisms proposed in this work.

\textbf{Exponential Probabilistic Growth (EPG) Model}: 
To ensure strict control of experimental variables and to directly validate the effectiveness of the exponential probabilistic growth mechanism proposed in this work, a BA-based modification termed the exponential probabilistic growth (EPG) network model is constructed. The EPG model preserves the structural framework of BA growth, where one node is added at each iteration and edges are attached through the standard preferential-attachment process. The only modification is that the fixed number of newly added edges is replaced by a random quantity generated in accordance with the exponential probabilistic growth formulation presented in Eq.~\ref{eq1}. Thus, this model can directly reflect the independent influence of the exponential probabilistic growth mechanism, which is one of the core mechanisms in the model proposed in this work. 

\textbf{Nonlinear Preferential Attachment (NPA) Model}: 
To ensure strict control of experimental variables and to directly examine the effectiveness of such attachment mechanisms, this study adopts a BA-based modification that introduces nonlinear preferential attachment (equal to the independent vari-linear preferential attachment). The resulting Nonlinear Preferential Attachment (NPA) model preserves the BA growth framework, where one node is added at each iteration and the number of newly added edges remains fixed at $m$. The only modification is that the attachment probability is adjusted from a linear dependence on node degree to a nonlinear form proportional to $k^{r}$. Where $k$ is the current degree and $r$ is a tunable exponent (following the definition in this work, also denoted as $\alpha$ in earlier research). Thus, this model can directly reflect the independent influence of the vari-linear preferential attachment mechanism, which is one of the core mechanisms in the model proposed in this work. 

\subsection{Optimal Fitting Calculation}\label{sec4-5}   

Similar to most other classical models, the vari-linear model is also a heuristic method. Unlike learning-based models, such heuristic methods do not rely on learning-based optimization and therefore require an independent procedure to determine their optimal parameters. This work employs a simple optimization method to highlight the intrinsic generative capability of the models rather than the performance of the fitting methods.

An important prerequisite for comparing such models is ensuring that the zeroth-order characteristics (node and edge counts) in each resulting network is approximately equivalent. The specific design of the computational scheme for the optimal network fitting results is as follows. 

Let the number of nodes and edges of the corresponding real network be $N$ and $M$, respectively. For the vari-linear model, the fixed inputs are the node counts $n=N$ and the expected average degree $k=2M/N$, and the variable is the vari-linear preferential attachment parameter $r$. 
For the ER model, the fixed inputs are the node counts $n=N$ and the connection probability $p=2M/N^2$.
For the WS model, the fixed inputs are the node counts $n=N$ and the neighbor node connection counts $k=2M/N$, reconnection probability $p=0.2$.
For the BA model, the fixed inputs are the node counts $n=N$ and the added edge counts $m=\text{round}(M/N)$ ($m\geq1$) for each step.
For the EPG model, the fixed inputs are the node counts $n=N$ and the expected value of added edge counts $k=2M/N$ for each step.
For the NPA model, the fixed inputs are the node counts $n$ and the added edge counts $m=\text{round}(M/N)$ ($m\geq1$) for each step, and the variable is the nonlinear preferential attachment parameter $\alpha$. 

During the optimization process, we use simple Bayesian optimization~\cite{RN709} to dynamically optimize parameters with the measurement results of each metrics as targets, ensuring scientific rigor in searching for the optimal network (one with the highest network similarity or the smallest degree distribution divergence).


\subsection{Multi-dimensional Evaluation Metrics}\label{sec4-6}

The metrics employed in this work to evaluate differences between networks are primarily divided into three metrics measuring network similarity (NLSD, SINS, and GDD) and three metrics measuring degree distribution divergence (WD, KS, and JS).


\textbf{Network Laplacian Spectral Descriptor (NLSD)}~\cite{RN790}: Compare graphs through a compact heat-trace signature derived from the normalized Laplacian spectrum. By encoding heat diffusion across multiple time scales, it captures both local and global structural patterns in a permutation-invariant manner. Its normalization strategy further ensures size-independent comparison, enabling efficient similarity measurement across heterogeneous empirical networks.

\textbf{Size-Independent Network Similarity (SINS)}~\cite{RN791}: Measures structural similarity between graphs of different sizes without requiring node correspondence. It constructs a compact ``signature'' vector for each network by aggregating local and egonet-based structural features into distributional moments, enabling comparisons that are both size-invariant and computationally efficient. This representation supports scalable similarity assessment across heterogeneous empirical networks.

\textbf{Graphlet Degree Distribution (GDD)}~\cite{RN792}: Characterize network structure by quantifying how nodes participate in small, non-isomorphic graphlets across their distinct automorphism-defined node roles. By constructing distributions over these graphlet-specific structural roles, GDD captures fine-grained local topology and imposes a rich set of constraints on network similarity, enabling sensitive comparison of networks through their agreement across graphlet-based patterns.


\textbf{Wasserstein Distance (WD)}~\cite{RN781}: Also known as kantorovich distance~\cite{RN780} or earth mover's distance~\cite{RN740}. Measures the minimum amount of work (moving cost) required to transform one distribution into another between two distributions. The underlying idea is to interpret each distribution as a pile of unit mass placed along the ordered degree axis. Transforming one distribution into another then amounts to shifting mass along this axis, where the cost of each shift is proportional to both the amount of mass moved and the distance it travels. The WD method is sensitive to divergences in the location and shape of distributions, and is widely employed in the comparison of graph structures~\cite{RN782}. 
In this work, the WD is computed using the \texttt{wasserstein\_distance} function from the \texttt{scipy.stats} package.

\textbf{Kolmogorov-Smirnov (KS) Test}~\cite{RN741, RN743}: It is a nonparametric test of the equality, it by identifying the divergence between their cumulative distribution functions. The KS test is sensitive to divergences in both location and shape of the empirical cumulative distribution functions of the two samples. And it can also serve as a goodness of fit test. The KS test is widely employed for comparing graph structures and quantifying network goodness of ~\cite{RN783, RN777}. 
In this work, the KS is computed using the \texttt{ks\_2samp} function from the \texttt{scipy.stats} package.

\textbf{Jensen-Shannon (JS) divergence}~\cite{RN739}: It is a symmetric measure of similarity between two probability distributions, based on the Kullback-Leibler (KL) divergence construction. It is constructed by comparing each distribution to their average reference distribution, which yields a smoothed and numerically stable assessment of how the two distributions differ. The JS divergence is widely employed to assess the similarity between real and generated networks~\cite{RN784}. 
In this work, the JS is computed using the \texttt{jensenshannon} function from the \texttt{scipy.spatial.distance} package.

\backmatter

\bmhead{Data availability}
This work utilizes real-world data sourced from SNAP (\href{https://snap.stanford.edu}{https://snap.stanford.edu}), KONECT (\href{http://konect.cc}{http://konect.cc}) and Compass (\href{https://www.scicompass.com}{https://www.scicompass.com}), the corresponding sources are listed in SI-Appendix~\ref{A-ds}.

\bmhead{Code availability}
The code necessary to reproduce the main plots and statistical analyses is available at GitHub (\href{https://github.com/Vinne-Git/Paper_RL}{https://github.com/Vinne-Git/Paper\_RL})

\bmhead{Acknowledgements}
This work is supported by the STI 2030-Major Projects (2022ZD0211400), the National Natural Science Foundation of China (T2293771), the Sichuan Province Outstanding Young Scientists Foundation (2023NSFSC1919) and the New Cornerstone Science Foundation through the XPLORER PRIZE.

\bmhead{Author contribution}
J.R. and L.L. conceptualized and designed the research, revised the manuscript and reviewed all the results. J.R. collected the experimental datasets, developed the algorithm, performed the numeric analysis and led in drafting the manuscript. We thank Xifei Fu for helpful discussions.

\bmhead{Competing interest declaration}
The authors declare no competing interests.


\bibliography{sn-article}

\clearpage
\begin{appendices}

\section*{Supplementary Information}

\section{Pseudocode of Vari-linear Network Generation Model}\label{A-code}

\begin{algorithm}[h]
\small
\caption{Vari-linear network generation model}\label{alg-model}
\KwData{
Total number of nodes $n$ ($n \in \mathbb{Z}^+$); expected average degree $k$ ($k > 2$); vari-linear parameter $r$ ($r \in \mathbb{R}$); initial graph size $m_0$ ($m_0 \in \mathbb{Z}^+$, $m_0 > 1$, default $m_0 = 3$).
}
\KwResult{Network $G$ generated by the vari-linear model.}
$G \gets$ CompleteGraph($m_0$)   \tcp*[r]{Initialize network}
\For{$i = m_0+1$ \KwTo $n$}{
    $m_i \sim P_g(\cdot)$   \tcp*[r]{ $P_g$ defined as Eq.~\eqref{eq2} }
    $m_i = \min(m_i, i)$ \;
    $w_j = {d_j}^r$ ($j \in G.Nodes$) \;
    $P_p(j) = w_j / \sum_{u} w_u$  \tcp*[r]{ $P_p$ defined as Eq.~\eqref{eq3} }
    RandomChoice($G.Nodes$, $m_i$, $P_p$)  \tcp*[r]{ $m_i$ nodes from $G$ selected with $P_p$ }
    \For{each $j \in $ RandomChoice($\cdot$)}{
        $G.AddEdge(i, j)$\;
    }
}
\Return{$G$}
\end{algorithm}

\clearpage
\section{Datasets}\label{A-ds}

\begin{table}[h]
\centering
\captionsetup{justification=raggedright,singlelinecheck=false} 
\caption{Real-World Network Datasets.}
\label{tab-data}
\begin{tabular}{m{2.8cm} m{1.3cm} m{2.4cm} m{1.2cm} m{1.2cm} m{1.2cm}}
\toprule
Network & Abbr. & Category & $N$ & $M$ & $\langle d \rangle$ \\
\midrule
RealityMining~\cite{RN717} & RM & Social & 96 & 2539 & 52.896 \\
Blogs~\cite{RN718} & BL & Social & 1224 & 16718 & 27.317 \\
IFM~\cite{RN719} & IM & Social & 1266 & 6451 & 10.191 \\
AmazonMTurk~\cite{RN765} & AMT & Social & 1389 & 5268 & 7.585 \\
SocHamsterster~\cite{RN719} & SH & Social & 2426 & 16630 & 13.710 \\
MovieLens~\cite{RN736} & ML & Social & 6040 & 987253 & 326.905 \\
Advogato~\cite{RN720} & AD & Social & 6539 & 43277 & 13.237 \\
LastFmAsia~\cite{RN721} & LF-A & Social & 7624 & 27806 & 7.294 \\
Facebook~\cite{RN430} & FB & Social & 11565 & 67114 & 11.606 \\
Brightkite~\cite{RN431} & BK & Social & 58228 & 214078 & 7.353 \\
Twitter~\cite{RN429} & X & Social & 81306 & 1768149 & 43.494 \\
CANetSci~\cite{RN722} & CA-NS & Co-Authorship & 1461 & 2742 & 3.754 \\
CAGrQc~\cite{RN719} & CA-GQ & Co-Authorship & 5242 & 14496 & 5.531 \\
CAAstroPh1~\cite{RN725} & CA-AP1 & Co-Authorship & 16046 & 121251 & 15.113 \\
CAAstroPh2~\cite{RN673} & CA-AP2 & Co-Authorship & 18772 & 198110 & 21.107 \\
CAConMat~\cite{RN724} & CA-CM & Co-Authorship & 30460 & 120029 & 7.881 \\
CitCora~\cite{RN725} & C-CR & Citation & 23166 & 89157 & 7.697 \\
CitHepPh2~\cite{RN673} & C-HP2 & Citation & 27770 & 352807 & 25.409 \\
CitHepPh1~\cite{RN428} & C-HP1 & Citation & 34546 & 421578 & 24.407 \\
EmailMC~\cite{RN727} & EM-MC & Communications & 167 & 3251 & 38.934 \\
EmailEC~\cite{RN673} & EM-EC & Communications & 1005 & 25571 & 50.888 \\
EmailRVU~\cite{RN728} & EM-RVU & Communications & 1133 & 5451 & 9.622 \\
Celegans~\cite{RN279} & CE & Biological & 297 & 2148 & 14.465 \\
Yeast~\cite{RN729} & YT & Biological & 1870 & 2277 & 2.435 \\
HumanProteins2~\cite{RN730} & HP-2 & Biological & 2239 & 6432 & 5.745 \\
HumanProteins1~\cite{RN731} & HP-1 & Biological & 3133 & 6726 & 4.294 \\
HsLc~\cite{RN732} & HL & Biological & 4227 & 39484 & 18.682 \\
Dmela~\cite{RN733} & DM & Biological & 7393 & 25569 & 6.917 \\
LesMis\'{e}rables~\cite{RN737} & LM & Literature$\&$Art & 77 & 254 & 6.597 \\
Jazz~\cite{RN734} & JZ & Literature$\&$Art & 198 & 2742 & 27.697 \\
Bible~\cite{RN736} & BB & Literature$\&$Art & 1773 & 9131 & 10.300 \\
Marvel~\cite{RN735} & MV & Literature$\&$Art & 19428 & 96662 & 9.951 \\
\bottomrule     
\end{tabular}
\begin{tablenotes}
    \footnotesize
    \item[*] Category denotes the real-world field the network belongs to. $N$ denotes the number of nodes in the network, $M$ denotes the number of connected edges in the network, and $\langle d \rangle$ denotes the average degree of the network.
\end{tablenotes}
\end{table}

\clearpage
\section{Numerical Results of Network Similarity}\label{A-ns}

\setcounter{table}{0} 

\begin{table}[h]
\centering
\resizebox{\linewidth}{!}{ 
\begin{minipage}{\textheight}
\centering
\footnotesize
\captionof{table}{Quantitative values of the similarity between each model and the corresponding real network based on the NLSD metric.}
\label{tab-ns-nlsd}
\begin{tabular}{
l >{\centering\arraybackslash}m{0.8cm} >{\centering\arraybackslash}m{0.8cm} >{\centering\arraybackslash}m{0.8cm} >{\centering\arraybackslash}m{0.8cm} >{\centering\arraybackslash}m{0.8cm} >{\centering\arraybackslash}m{0.8cm} >{\centering\arraybackslash}m{0.8cm}
}
\toprule
Networks & ER & WS & BA & DG & GGDP & GW & Our \\
\midrule
LesMis\'{e}rables & 0.1789 & 0.1781 & 0.1748 & 0.2033 & 0.3585 & 2.8612 & 0.0831 \\
RealityMining & 0.0095 & 0.0090 & 0.0053 & 0.0101 & 0.0021 & 0.1172 & 0.0012 \\
EmailMC & 0.0273 & 0.0190 & 0.0219 & 0.0305 & 0.0438 & 0.4535 & 0.0007 \\
Jazz & 0.0751 & 0.0613 & 0.0736 & 0.0779 & 0.1001 & 2.1070 & 0.0042 \\
Celegans & 0.0270 & 0.0124 & 0.0295 & 0.0247 & 0.0674 & 0.3157 & 0.0010 \\
EmailEC & 0.0955 & 0.0911 & 0.0954 & 0.0943 & 0.0982 & 0.0989 & 0.0792 \\
EmailRVU & 0.0166 & 0.0121 & 0.0179 & 0.0139 & 0.0326 & 8.8113 & 0.0008 \\
Blogs & 0.0231 & 0.0175 & 0.0233 & 0.0219 & 0.0293 & 0.0880 & 0.0058 \\
IFM & 0.0128 & 0.0080 & 0.0130 & 0.0093 & 0.0267 & 9.8760 & 0.0005 \\
AmazonMTurk & 0.0479 & 0.0604 & 0.0631 & 0.0151 & 0.0711 & 8.0100 & 0.0513 \\
CANetSci & 0.0484 & 0.0500 & 0.0536 & 0.0438 & 0.0676 & 7.7975 & 0.0362 \\
Bible & 0.0368 & 0.0483 & 0.0504 & 0.0448 & 0.0557 & - & 0.0411 \\
Yeast & 0.0115 & 0.0006 & 0.0046 & 0.0265 & - & - & 0.0089 \\
HumanProteins2 & 0.0338 & 0.0329 & 0.0349 & 0.0241 & 0.0416 & 0.7526 & 0.0278 \\
SocHamsterster & 0.0376 & 0.0319 & 0.0377 & 0.0369 & - & 3.3236 & 0.0310 \\
HumanProteins1 & 0.0208 & 0.0229 & 0.0214 & 0.0120 & - & - & 0.0178 \\
HsLc & 0.0154 & 0.0142 & 0.0154 & - & - & 9.0646 & 0.0106 \\
CAGrQc & 0.0129 & 0.0152 & 0.0161 & - & - & - & 0.0132 \\
MovieLens & 0.0007 & 0.0012 & 0.0007 & 0.0008 & - & 0.0124 & 0.0003 \\
Advogato & 0.0139 & 0.0133 & 0.0140 & - & - & 0.6907 & 0.0112 \\
Dmela & 0.0040 & 0.0036 & 0.0039 & - & - & - & 0.0012 \\
LastFmAsia & 0.0069 & 0.0064 & 0.0072 & - & - & - & 0.0042 \\
Facebook & 0.0056 & 0.0051 & 0.0056 & - & - & - & 0.0037 \\
CAAstroPh1 & 0.0067 & 0.0052 & 0.0055 & - & - & - & 0.0043 \\
CAAstroPh2 & 0.0050 & 0.0047 & 0.0045 & - & - & - & 0.0040 \\
Marvel & 0.0037 & 0.0035 & 0.0046 & - & - & 7.8981 & 0.0028 \\
CitCora & 0.0026 & 0.0024 & 0.0027 & - & - & - & 0.0018 \\
CitHepPh2 & 0.0034 & 0.0028 & 0.0030 & - & - & - & 0.0028 \\
CAConMat & 0.0026 & 0.0025 & 0.0027 & - & - & - & 0.0021 \\
CitHepPh1 & 0.0027 & 0.0026 & 0.0023 & - & - & - & 0.0022 \\
Brightkite & 0.0013 & 0.0013 & 0.0015 & - & - & - & 0.0012 \\
Twitter & 0.0010 & 0.0009 & 0.0010 & - & - & - & 0.0007 \\
\bottomrule
\end{tabular}
\begin{tablenotes}
    \footnotesize
    \item[*] The NLSD metric denotes the network Laplacian spectral descriptor, which is a positive-valued metric, meaning that lower values indicating higher similarity between networks.
    \item[*] The symbol `-' denotes cases where learning-based models are unable to perform computations even under the conditions of our high-performance computing platform, particularly for large networks.
\end{tablenotes}
\end{minipage}
} 
\end{table}

\clearpage

\vfill
\begin{table}[h]
\centering
\resizebox{\linewidth}{!}{ 
\begin{minipage}{\textheight}
\centering
\footnotesize
\captionof{table}{Quantitative values of the similarity between each model and the corresponding real network based on the SINS metric.}
\label{tab-ns-sins}
\begin{tabular}{
l >{\centering\arraybackslash}m{0.8cm} >{\centering\arraybackslash}m{0.8cm} >{\centering\arraybackslash}m{0.8cm} >{\centering\arraybackslash}m{0.8cm} >{\centering\arraybackslash}m{0.8cm} >{\centering\arraybackslash}m{0.8cm} >{\centering\arraybackslash}m{0.8cm}
}
\toprule
Networks & ER & WS & BA & DG & GGDP & GW & Our \\
\midrule
LesMis\'{e}rables & 16.578 & 17.510 & 16.039 & 12.590 & 21.793 & 19.583 & 12.596 \\
RealityMining & 17.048 & 15.081 & 14.024 & 10.657 & 19.797 & 23.234 & 7.532 \\
EmailMC & 19.892 & 22.110 & 15.220 & 8.973 & 19.788 & 28.897 & 7.937 \\
Jazz & 19.699 & 19.799 & 15.596 & 10.998 & 23.172 & 22.792 & 7.417 \\
Celegans & 21.044 & 18.972 & 12.034 & 22.214 & 26.004 & 22.550 & 8.607 \\
EmailEC & 23.484 & 21.474 & 12.999 & 10.438 & 27.774 & 20.862 & 10.186 \\
EmailRVU & 21.938 & 16.563 & 13.267 & 10.748 & 29.592 & 24.990 & 10.719 \\
Blogs & 24.001 & 21.797 & 15.557 & 9.613 & 28.869 & 21.012 & 9.890 \\
IFM & 22.966 & 20.502 & 12.419 & 7.821 & 30.850 & 30.095 & 3.663 \\
AmazonMTurk & 24.978 & 23.984 & 18.344 & 13.990 & 29.807 & 27.831 & 13.067 \\
CANetSci & 21.829 & 22.821 & 22.521 & 18.896 & 28.709 & 30.061 & 16.124 \\
Bible & 25.485 & 24.411 & 17.454 & 13.443 & 29.263 & - & 11.535 \\
Yeast & 18.795 & 23.434 & 16.433 & 11.653 & - & - & 11.438 \\
HumanProteins2 & 23.315 & 24.064 & 13.446 & 11.730 & 31.462 & 17.567 & 8.722 \\
SocHamsterster & 26.208 & 24.022 & 17.265 & 13.313 & - & 25.823 & 13.800 \\
HumanProteins1 & 21.888 & 17.029 & 11.477 & 10.039 & - & - & 16.239 \\
HsLc & 24.762 & 22.808 & 15.904 & - & - & 32.085 & 15.302 \\
CAGrQc & 23.671 & 23.322 & 20.916 & - & - & - & 20.935 \\
MovieLens & 26.292 & 22.608 & 14.405 & 6.930 & - & 21.364 & 7.481 \\
Advogato & 26.058 & 22.793 & 17.214 & - & - & 25.371 & 17.552 \\
Dmela & 22.595 & 24.008 & 11.071 & - & - & - & 5.230 \\
LastFmAsia & 25.585 & 20.343 & 17.525 & - & - & - & 15.699 \\
Facebook & 26.232 & 21.641 & 17.318 & - & - & - & 16.320 \\
CAAstroPh1 & 26.576 & 24.956 & 20.338 & - & - & - & 20.170 \\
CAAstroPh2 & 26.560 & 24.673 & 19.698 & - & - & - & 19.276 \\
Marvel & 27.853 & 31.215 & 22.680 & - & - & 19.342 & 16.286 \\
CitCora & 25.359 & 20.702 & 18.439 & - & - & - & 17.057 \\
CitHepPh2 & 26.821 & 22.681 & 18.246 & - & - & - & 17.882 \\
CAConMat & 24.904 & 24.636 & 19.861 & - & - & - & 16.859 \\
CitHepPh1 & 25.506 & 22.311 & 17.742 & - & - & - & 15.661 \\
Brightkite & 25.614 & 23.892 & 18.160 & - & - & - & 15.398 \\
Twitter & 28.631 & 25.948 & 19.912 & - & - & - & 22.706 \\
\bottomrule
\end{tabular}
\begin{tablenotes}
    \footnotesize
    \item[*] The SINS metric denotes the size-independent network similarity, which is a positive-valued metric, meaning that lower values indicating higher similarity between networks. 
    \item[*] The symbol `-' denotes cases where learning-based models are unable to perform computations even under the conditions of our high-performance computing platform, particularly for large networks.
\end{tablenotes}
\end{minipage}
} 
\end{table}
\vfill

\clearpage

\begin{table}[h]
\centering
\resizebox{\linewidth}{!}{ 
\begin{minipage}{\textheight}
\centering
\footnotesize
\captionof{table}{Quantitative values of the similarity between each model and the corresponding real network based on the GDD metric.}
\label{tab-ns-gdd}
\begin{tabular}{
l >{\centering\arraybackslash}m{0.8cm} >{\centering\arraybackslash}m{0.8cm} >{\centering\arraybackslash}m{0.8cm} >{\centering\arraybackslash}m{0.8cm} >{\centering\arraybackslash}m{0.8cm} >{\centering\arraybackslash}m{0.8cm} >{\centering\arraybackslash}m{0.8cm}
}
\toprule
Networks & ER & WS & BA & DG & GGDP & GW & Our \\
\midrule
LesMis\'{e}rables & 0.4286 & 0.3052 & 0.6494 & 0.7805 & 0.4221 & 0.6249 & 0.7273 \\
RealityMining & 0.6927 & 0.6719 & 0.5104 & 0.6979 & 0.6406 & 0.2877 & 0.7292 \\
EmailMC & 0.5539 & 0.4611 & 0.6168 & 0.7515 & 0.5868 & 0.3314 & 0.7784 \\
Jazz & 0.5758 & 0.4369 & 0.6364 & 0.6878 & 0.3914 & 0.5554 & 0.8460 \\
Celegans & 0.2205 & 0.2121 & 0.6380 & 0.3855 & 0.1498 & 0.4171 & 0.7290 \\
EmailEC & 0.1652 & 0.1483 & 0.8910 & 0.8465 & 0.1378 & 0.6076 & 0.9303 \\
EmailRVU & 0.4841 & 0.2917 & 0.7056 & 0.9327 & 0.1717 & 0.5437 & 0.8892 \\
Blogs & 0.1703 & 0.1622 & 0.9040 & 0.8714 & 0.1054 & 0.7228 & 0.9109 \\
IFM & 0.4491 & 0.1781 & 0.8780 & 0.9501 & 0.0762 & 0.7160 & 0.9491 \\
AmazonMTurk & 0.4633 & 0.1760 & 0.9374 & 0.8812 & 0.0274 & 0.8206 & 0.9352 \\
CANetSci & 0.5913 & 0.4535 & 0.6872 & 0.7850 & 0.0431 & 0.6290 & 0.8908 \\
Bible & 0.4185 & 0.0778 & 0.9464 & 0.9443 & 0.0341 & - & 0.9780 \\
Yeast & 0.5664 & 0.5874 & 0.8658 & 0.8395 & - & - & 0.8481 \\
HumanProteins2 & 0.4975 & 0.2307 & 0.9580 & 0.9831 & 0.0226 & 0.9698 & 0.9866 \\
SocHamsterster & 0.3636 & 0.1105 & 0.8706 & 0.9509 & - & 0.8272 & 0.9518 \\
HumanProteins1 & 0.5283 & 0.4177 & 0.9167 & 0.9398 & - & - & 0.8886 \\
HsLc & 0.2935 & 0.0882 & 0.9247 & - & - & 0.8489 & 0.8608 \\
CAGrQc & 0.5889 & 0.2689 & 0.8409 & - & - & - & 0.8136 \\
MovieLens & 0.1255 & 0.1077 & 0.9493 & 0.9546 & - & 0.6848 & 0.9600 \\
Advogato & 0.4365 & 0.0525 & 0.9831 & - & - & 0.9464 & 0.9493 \\
Dmela & 0.5465 & 0.1508 & 0.9200 & - & - & - & 0.8998 \\
LastFmAsia & 0.5245 & 0.1552 & 0.9370 & - & - & - & 0.8970 \\
Facebook & 0.5039 & 0.1086 & 0.9300 & - & - & - & 0.9027 \\
CAAstroPh1 & 0.4600 & 0.0955 & 0.8307 & - & - & - & 0.9285 \\
CAAstroPh2 & 0.4209 & 0.0773 & 0.8531 & - & - & - & 0.9373 \\
Marvel & 0.4937 & 0.1258 & 0.9822 & - & - & 0.9944 & 0.9977 \\
CitCora & 0.5050 & 0.1594 & 0.9518 & - & - & - & 0.9356 \\
CitHepPh2 & 0.3798 & 0.0109 & 0.9850 & - & - & - & 0.9863 \\
CAConMat & 0.5286 & 0.1643 & 0.8939 & - & - & - & 0.9664 \\
CitHepPh1 & 0.4180 & 0.0368 & 0.9227 & - & - & - & 0.9074 \\
Brightkite & 0.5072 & 0.1122 & 0.9950 & - & - & - & 0.9877 \\
Twitter & 0.3077 & 0.0120 & 0.9904 & - & - & - & 0.9793 \\
\bottomrule
\end{tabular}
\begin{tablenotes}
    \footnotesize
    \item[*] The GDD metric denotes the graphlet degree distribution, which is a bounded metric in $[0,1]$, meaning that higher values indicating greater similarity between networks.
    \item[*] The symbol `-' denotes cases where learning-based models are unable to perform computations even under the conditions of our high-performance computing platform, particularly for large networks.
\end{tablenotes}
\end{minipage}
} 
\end{table}

\clearpage
\section{Degree Distribution Comparing Vari-linear Network Versus Real-World Networks}\label{A-rw-dd}

\begin{figure}[h]
\centering
\includegraphics[width=\textwidth, height=0.62\textheight, keepaspectratio]{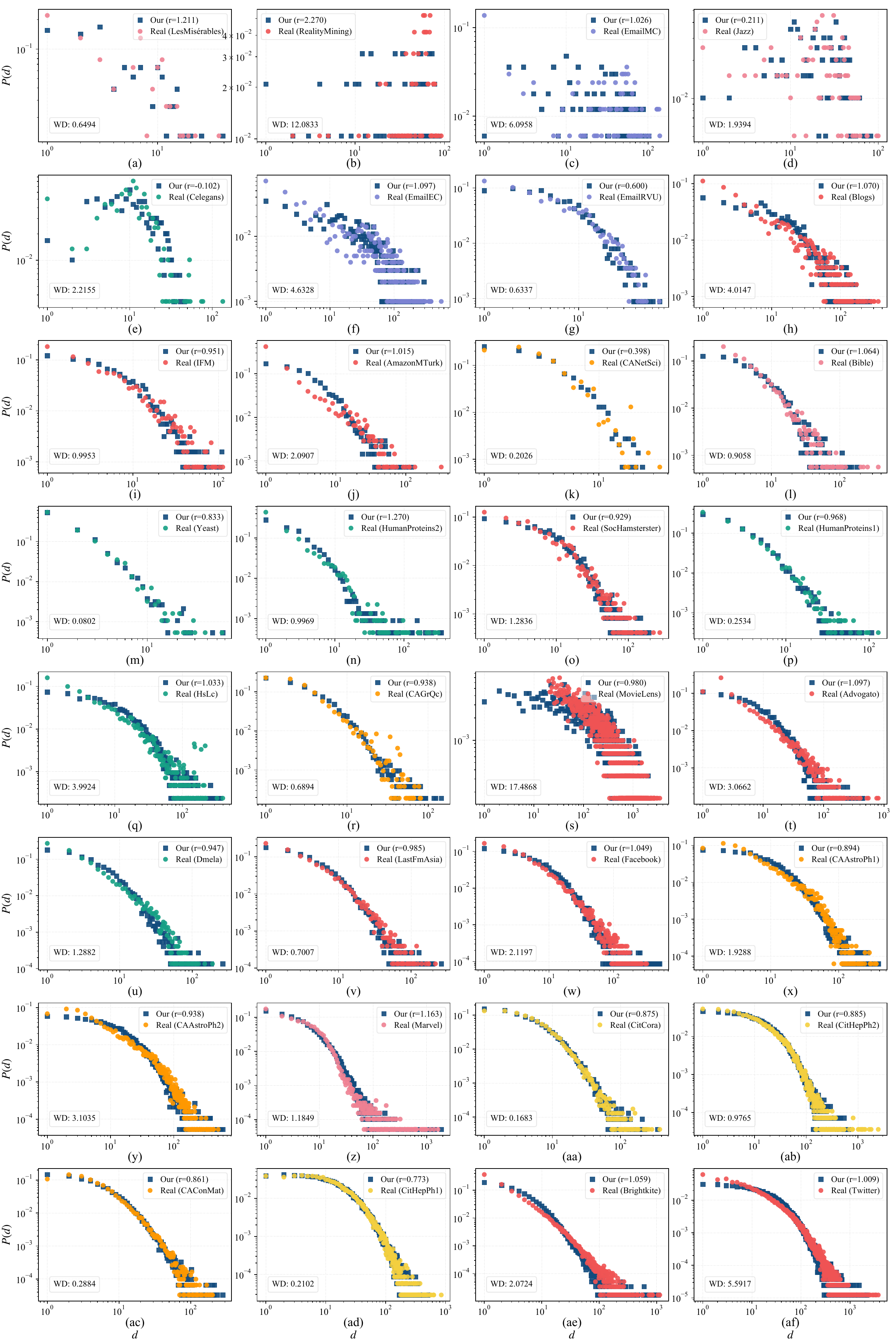}
\caption{Comparison of degree distribution probabilities (log-log scale) between the optimal result network of vari-linear model and the real network (with WD metric as the optimization objective). In all subfigures, the blue scatters represent the degree distribution of vari-linear network in the optimal case, and the scatters in other colors are the degree distributions of the corresponding real-world networks. Categories of real-world networks include: social networks (red), scholarly co-authorship networks (orange), academic citation networks (yellow), communication networks (purple), biological networks (green) and literary \& artistic networks (pink).}
\label{fig-dd-wd} 
\end{figure}

\begin{figure}[h]
\centering
\includegraphics[width=\textwidth, height=0.75\textheight, keepaspectratio]{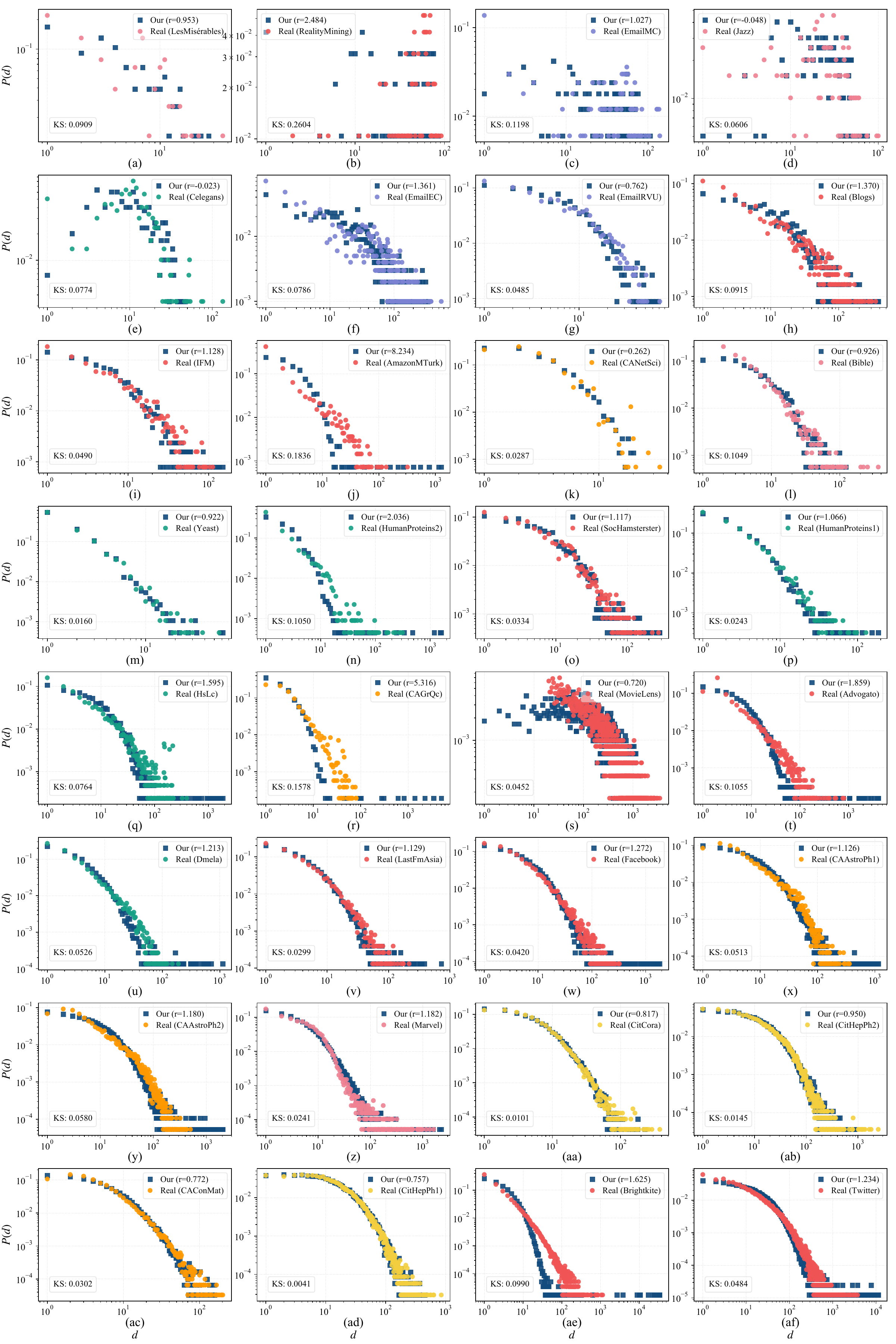}
\caption{
Comparison of degree distribution probabilities (log-log scale) between the optimal result network of vari-linear model and the real network (with KS metric as the optimization objective). In all subfigures, the blue scatters represent the degree distribution of vari-linear network in the optimal case, and the scatters in other colors are the degree distributions of the corresponding real-world networks. Categories of real-world networks include: social networks (red), scholarly co-authorship networks (orange), academic citation networks (yellow), communication networks (purple), biological networks (green) and literary \& artistic networks (pink).
}
\label{fig-dd-ks} 
\end{figure}

\begin{figure}[h]
\centering
\includegraphics[width=\textwidth, height=0.75\textheight, keepaspectratio]{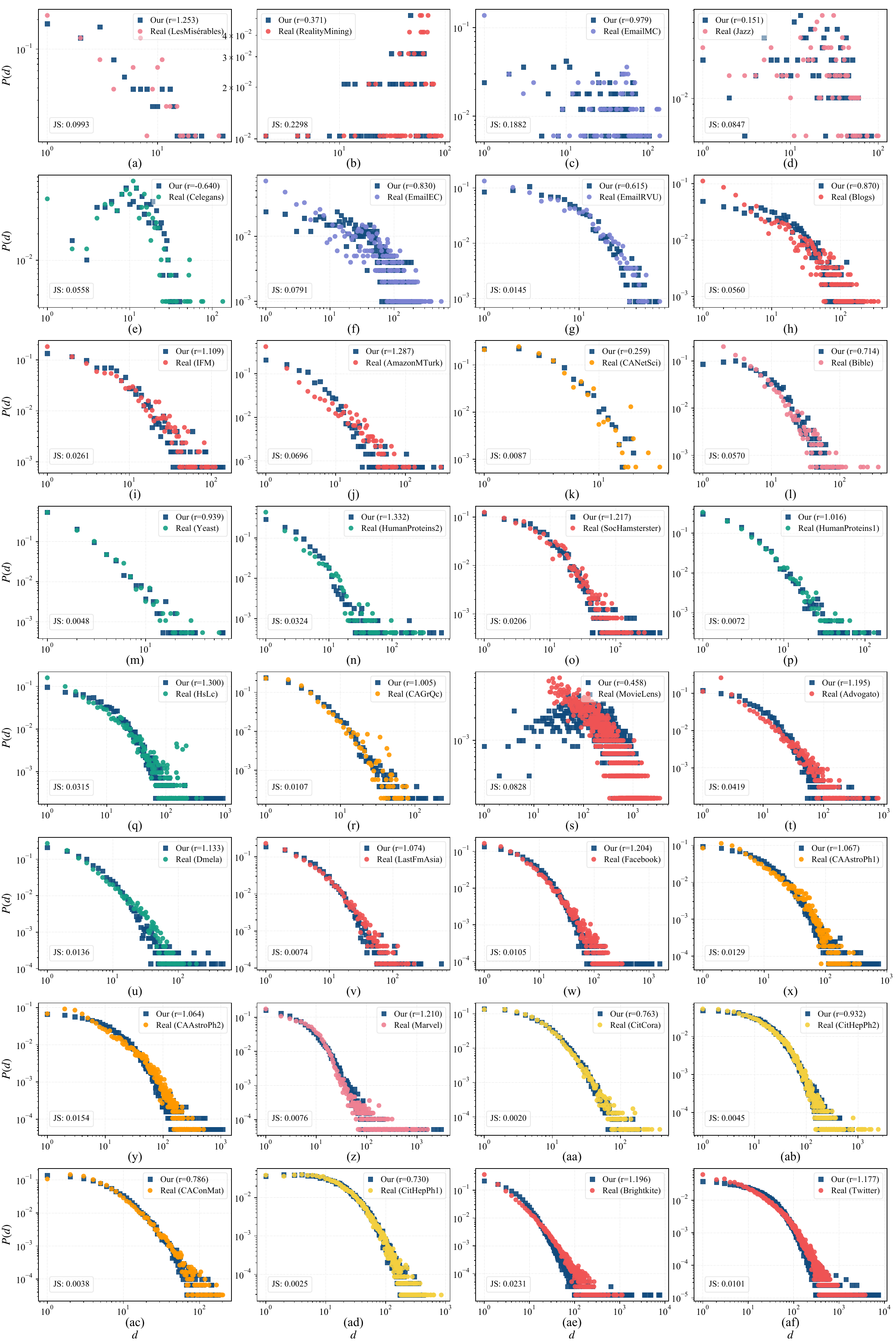}
\caption{Comparison of degree distribution probabilities (log-log scale) between the optimal result network of vari-linear model and the real network (with JS metric as the optimization objective). In all subfigures, the blue scatters represent the degree distribution of vari-linear network in the optimal case, and the scatters in other colors are the degree distributions of the corresponding real-world networks. Categories of real-world networks include: social networks (red), scholarly co-authorship networks (orange), academic citation networks (yellow), communication networks (purple), biological networks (green) and literary \& artistic networks (pink).}
\label{fig-dd-js} 
\end{figure}

\clearpage
\section{Statistical Assessment of Result Disparities}\label{A-sa}

\setcounter{table}{0} 

\begin{table}[h]
\centering
\captionsetup{justification=raggedright,singlelinecheck=false} 
\rotatebox{90}{ 
\resizebox{0.8\textheight}{!}{ 
\begin{minipage}{\textheight}  
\centering
\captionof{table}{Statistical analysis of the results advantage of the vari-linear model based on all empirical datasets and six metrics.}
\label{tab-sa}
\begin{tabular}{cc}
\toprule
\begin{tabular}{lcccc}
Metric & Baseline & Ratio & 95\% CI & $p$ \\
\midrule
\multirow{6}{*}{NLSD} 
& ER    & 2.35$\times$ & [1.63, 3.52] & $2.807\times 10^{-5}$ \\
& WS    & 1.97$\times$ & [1.32, 2.97] & $7.198\times 10^{-6}$ \\
& BA    & 2.28$\times$ & [1.60, 3.40] & $4.226\times 10^{-6}$ \\
& DG    & 3.41$\times$ & [1.75, 6.84] & $7.904\times 10^{-3}$ \\
& GGDP  & 6.66$\times$ & [2.81, 16.69] & $2.441\times 10^{-4}$ \\
& GW    & 193.84$\times$ & [64.81, 570.87] & $1.526\times 10^{-5}$ \\
\midrule
\multirow{6}{*}{SINS} 
& ER    & 1.90$\times$ & [1.68, 2.17] & $7.953\times 10^{-7}$ \\
& WS    & 1.77$\times$ & [1.56, 2.04] & $7.953\times 10^{-7}$ \\
& BA    & 1.32$\times$ & [1.19, 1.47] & $2.189\times 10^{-5}$ \\
& DG    & 1.17$\times$ & [1.02, 1.37] & $5.054\times 10^{-2}$ \\
& GGDP  & 2.83$\times$ & [2.36, 3.51] & $2.441\times 10^{-4}$ \\
& GW    & 2.33$\times$ & [1.92, 2.87] & $1.526\times 10^{-5}$ \\
\midrule
\multirow{6}{*}{GDD} 
& ER    & 2.16$\times$ & [1.89, 2.51] & $7.953\times 10^{-7}$ \\
& WS    & 6.15$\times$ & [4.34, 8.88] & $7.953\times 10^{-7}$ \\
& BA    & 1.06$\times$ & [1.02, 1.10] & $2.484\times 10^{-2}$ \\
& DG    & 1.07$\times$ & [1.01, 1.17] & $5.054\times 10^{-2}$ \\
& GGDP  & 6.97$\times$ & [3.73, 13.27] & $2.441\times 10^{-4}$ \\
& GW    & 1.37$\times$ & [1.21, 1.57] & $1.526\times 10^{-5}$ \\
\end{tabular}
&
\begin{tabular}{lcccc}
Metric & Baseline & Ratio & 95\% CI & $p$ \\
\midrule
\multirow{6}{*}{WD} 
& ER    & 6.23$\times$ & [4.77, 8.10] & $7.953\times 10^{-7}$ \\
& WS    & 7.07$\times$ & [5.39, 9.19] & $7.953\times 10^{-7}$ \\
& BA    & 3.60$\times$ & [2.89, 4.50] & $7.953\times 10^{-7}$ \\
& DG    & 2.33$\times$ & [1.53, 3.66] & $6.653\times 10^{-3}$ \\
& GGDP  & 135.17$\times$ & [40.68, 425.69] & $2.441\times 10^{-4}$ \\
& GW    & 6.85$\times$ & [5.37, 8.71] & $1.526\times 10^{-5}$ \\
\midrule
\multirow{6}{*}{KS} 
& ER    & 8.67$\times$ & [6.43, 11.88] & $7.948\times 10^{-7}$ \\
& WS    & 10.36$\times$ & [7.63, 14.08] & $7.953\times 10^{-7}$ \\
& BA    & 8.06$\times$ & [6.01, 10.90] & $7.953\times 10^{-7}$ \\
& DG    & 2.14$\times$ & [1.52, 3.14] & $2.579\times 10^{-3}$ \\
& GGDP  & 11.00$\times$ & [7.57, 15.59] & $2.441\times 10^{-4}$ \\
& GW    & 9.61$\times$ & [6.77, 13.45] & $1.526\times 10^{-5}$ \\
\midrule
\multirow{6}{*}{JS} 
& ER    & 12.58$\times$ & [8.48, 18.18] & $7.953\times 10^{-7}$ \\
& WS    & 17.83$\times$ & [11.83, 26.45] & $7.953\times 10^{-7}$ \\
& BA    & 11.32$\times$ & [7.69, 16.63] & $7.953\times 10^{-7}$ \\
& DG    & 1.69$\times$ & [1.35, 2.25] & $4.578\times 10^{-5}$ \\
& GGDP  & 12.00$\times$ & [7.00, 20.96] & $2.441\times 10^{-4}$ \\
& GW    & 8.63$\times$ & [5.33, 13.86] & $1.526\times 10^{-5}$ \\
\end{tabular} \\
\bottomrule
\end{tabular}
\begin{tablenotes}
    \footnotesize
    \item[*] Ratio: geometric-mean performance ratio, values $>1$ indicate better average performance.
    \item[*] 95\% CI: bootstrap confidence interval, lower bound $>1$ indicates a robust advantage.
    \item[*] $p$: Wilcoxon signed-rank test, smaller values indicate more reliable advantages (generally, $p<0.05$ reliable, $p<10^{-2}$ strong, $p<10^{-3}$ very strong).
\end{tablenotes}
\end{minipage}
}} 
\end{table}

\clearpage
\section{Numerical Results of Degree Distribution Divergence}\label{A-dm-dd}

\setcounter{table}{0} 

\begin{table}[h]
\centering
\resizebox{\linewidth}{!}{ 
\begin{minipage}{\textheight}
\centering
\footnotesize
\captionof{table}{Quantitative values of degree distribution divergences between each model and the corresponding real network based on the WD metric.}
\label{tab-dd-wd}
\begin{tabular}{
l >{\centering\arraybackslash}m{0.8cm} >{\centering\arraybackslash}m{0.8cm} >{\centering\arraybackslash}m{0.8cm} >{\centering\arraybackslash}m{0.8cm} >{\centering\arraybackslash}m{0.9cm} >{\centering\arraybackslash}m{0.8cm} >{\centering\arraybackslash}m{0.8cm}
}
\toprule
Networks & ER & WS & BA & DG & GGDP & GW & Our \\
\midrule
LesMis\'{e}rables & 2.70 & 3.61 & 2.23 & 0.73 & 28.26 & 3.88 & 0.65 \\
RealityMining & 12.12 & 12.23 & 16.08 & 2.94 & 16.58 & 43.95 & 12.08 \\
EmailMC & 21.15 & 22.12 & 14.98 & 4.96 & 41.76 & 35.05 & 6.10 \\
Jazz & 9.83 & 11.39 & 5.48 & 2.32 & 60.77 & 23.35 & 1.94 \\
Celegans & 4.92 & 5.73 & 2.69 & 4.68 & 125.40 & 12.76 & 2.22 \\
EmailEC & 37.20 & 39.56 & 20.05 & 19.86 & 432.57 & 46.39 & 4.63 \\
EmailRVU & 4.56 & 5.69 & 3.06 & 2.25 & 515.19 & 7.80 & 0.63 \\
Blogs & 22.43 & 23.99 & 12.58 & 13.58 & 578.73 & 23.83 & 4.01 \\
IFM & 6.56 & 7.38 & 3.52 & 1.32 & 598.13 & 8.99 & 1.00 \\
AmazonMTurk & 6.34 & 7.27 & 4.02 & 3.72 & 675.61 & 6.09 & 2.09 \\
CANetSci & 1.00 & 1.56 & 0.86 & 1.14 & 716.07 & 2.75 & 0.20 \\
Bible & 6.52 & 7.23 & 3.13 & 5.94 & 810.95 & - & 0.91 \\
Yeast & 0.96 & 0.90 & 0.54 & 1.26 & - & - & 0.08 \\
HumanProteins2 & 4.75 & 5.32 & 2.82 & 2.24 & 1027.56 & 3.76 & 1.00 \\
SocHamsterster & 9.47 & 10.63 & 5.05 & 3.96 & - & 11.78 & 1.28 \\
HumanProteins1 & 2.28 & 2.55 & 1.08 & 1.08 & - & - & 0.25 \\
HsLc & 16.12 & 17.86 & 9.33 & - & - & 17.18 & 3.99 \\
CAGrQc & 3.22 & 4.02 & 1.84 & - & - & - & 0.69 \\
MovieLens & 246.00 & 250.87 & 113.68 & 26.14 & - & 297.65 & 17.49 \\
Advogato & 11.87 & 13.08 & 7.10 & - & - & 11.26 & 3.07 \\
Dmela & 4.82 & 5.98 & 2.90 & - & - & - & 1.29 \\
LastFmAsia & 4.61 & 5.61 & 2.38 & - & - & - & 0.70 \\
Facebook & 9.19 & 10.15 & 5.09 & - & - & - & 2.12 \\
CAAstroPh1 & 11.01 & 12.38 & 6.28 & - & - & - & 1.93 \\
CAAstroPh2 & 16.50 & 18.10 & 9.33 & - & - & - & 3.10 \\
Marvel & 7.03 & 7.69 & 3.49 & - & - & 8.95 & 1.18 \\
CitCora & 3.96 & 4.81 & 1.77 & - & - & - & 0.17 \\
CitHepPh2 & 17.77 & 19.40 & 8.48 & - & - & - & 0.98 \\
CAConMat & 4.19 & 4.95 & 2.01 & - & - & - & 0.29 \\
CitHepPh1 & 15.20 & 16.41 & 7.42 & - & - & - & 0.21 \\
Brightkite & 6.39 & 7.31 & 3.92 & - & - & - & 2.07 \\
Twitter & 37.96 & 39.82 & 20.21 & - & - & - & 5.59 \\
\bottomrule
\end{tabular}
\begin{tablenotes}
    \footnotesize
    \item[*] The WD metric denotes the wasserstein distance.
    \item[*] The symbol `-' denotes cases where learning-based models are unable to perform computations even under the conditions of our high-performance computing platform, particularly for large networks.
\end{tablenotes}
\end{minipage}
} 
\end{table}

\clearpage

\begin{table}[h]
\centering
\resizebox{\linewidth}{!}{ 
\begin{minipage}{\textheight}
\centering
\footnotesize
\captionof{table}{Quantitative values of degree distribution divergences between each model and the corresponding real network based on the KS metric.}
\label{tab-dd-ks}
\begin{tabular}{
l >{\centering\arraybackslash}m{0.8cm} >{\centering\arraybackslash}m{0.8cm} >{\centering\arraybackslash}m{0.8cm} >{\centering\arraybackslash}m{0.8cm} >{\centering\arraybackslash}m{0.9cm} >{\centering\arraybackslash}m{0.8cm} >{\centering\arraybackslash}m{0.8cm}
}
\toprule
Networks & ER & WS & BA & DG & GGDP & GW & Our \\
\midrule
LesMis\'{e}rables & 0.3636 & 0.4545 & 0.3506 & 0.1081 & 0.9870 & 0.4525 & 0.0909 \\
RealityMining & 0.4271 & 0.3646 & 0.5000 & 0.1354 & 0.5729 & 0.8637 & 0.2604 \\
EmailMC & 0.4371 & 0.4491 & 0.2934 & 0.1677 & 0.8263 & 0.7144 & 0.1198 \\
Jazz & 0.3333 & 0.3889 & 0.2222 & 0.0602 & 0.9798 & 0.7384 & 0.0606 \\
Celegans & 0.2525 & 0.3502 & 0.1953 & 0.1852 & 0.9966 & 0.8760 & 0.0774 \\
EmailEC & 0.5204 & 0.5741 & 0.4219 & 0.2046 & 0.9990 & 0.6852 & 0.0786 \\
EmailRVU & 0.3372 & 0.4766 & 0.3751 & 0.0941 & 1.0000 & 0.7758 & 0.0485 \\
Blogs & 0.5703 & 0.6225 & 0.5131 & 0.1515 & 1.0000 & 0.6331 & 0.0915 \\
IFM & 0.4297 & 0.5300 & 0.4439 & 0.1154 & 1.0000 & 0.7167 & 0.0490 \\
AmazonMTurk & 0.5601 & 0.6523 & 0.6192 & 0.2783 & 1.0000 & 0.3808 & 0.1836 \\
CANetSci & 0.1808 & 0.3539 & 0.2101 & 0.1320 & 1.0000 & 0.7899 & 0.0287 \\
Bible & 0.4664 & 0.5713 & 0.4473 & 0.1388 & 1.0000 & - & 0.1049 \\
Yeast & 0.3009 & 0.2337 & 0.1139 & 0.2689 & - & - & 0.0160 \\
HumanProteins2 & 0.5205 & 0.6521 & 0.5873 & 0.2086 & 1.0000 & 0.5254 & 0.1050 \\
SocHamsterster & 0.4946 & 0.5816 & 0.4827 & 0.1221 & - & 0.7747 & 0.0334 \\
HumanProteins1 & 0.3585 & 0.4437 & 0.3367 & 0.1417 & - & - & 0.0243 \\
HsLc & 0.5815 & 0.6563 & 0.5392 & - & - & 0.6155 & 0.0764 \\
CAGrQc & 0.3919 & 0.5616 & 0.4407 & - & - & - & 0.1578 \\
MovieLens & 0.6139 & 0.6334 & 0.4379 & 0.0710 & - & 0.7917 & 0.0452 \\
Advogato & 0.5952 & 0.6756 & 0.6091 & - & - & 0.8741 & 0.1055 \\
Dmela & 0.4678 & 0.6238 & 0.4454 & - & - & - & 0.0526 \\
LastFmAsia & 0.4203 & 0.5728 & 0.4896 & - & - & - & 0.0299 \\
Facebook & 0.5318 & 0.6295 & 0.5412 & - & - & - & 0.0420 \\
CAAstroPh1 & 0.5201 & 0.6075 & 0.5191 & - & - & - & 0.0513 \\
CAAstroPh2 & 0.5511 & 0.6169 & 0.5283 & - & - & - & 0.0580 \\
Marvel & 0.4577 & 0.5645 & 0.4490 & - & - & 0.8254 & 0.0241 \\
CitCora & 0.3617 & 0.5001 & 0.3840 & - & - & - & 0.0101 \\
CitHepPh2 & 0.5146 & 0.5903 & 0.4588 & - & - & - & 0.0145 \\
CAConMat & 0.3766 & 0.5061 & 0.3820 & - & - & - & 0.0302 \\
CitHepPh1 & 0.4812 & 0.5443 & 0.3979 & - & - & - & 0.0041 \\
Brightkite & 0.5535 & 0.6722 & 0.6164 & - & - & - & 0.0990 \\
Twitter & 0.6096 & 0.6565 & 0.5380 & - & - & - & 0.0484 \\
\bottomrule
\end{tabular}
\begin{tablenotes}
    \footnotesize
    \item[*] The KS metric denotes the Kolmogorov-Smirnov test.
    \item[*] The symbol `-' denotes cases where learning-based models are unable to perform computations even under the conditions of our high-performance computing platform, particularly for large networks.
\end{tablenotes}
\end{minipage}
} 
\end{table}

\clearpage

\begin{table}[h]
\centering
\resizebox{\linewidth}{!}{ 
\begin{minipage}{\textheight}
\centering
\footnotesize
\captionof{table}{Quantitative values of degree distribution divergences between each model and the corresponding real network based on the JS metric.}
\label{tab-dd-js}
\begin{tabular}{
l >{\centering\arraybackslash}m{0.8cm} >{\centering\arraybackslash}m{0.8cm} >{\centering\arraybackslash}m{0.8cm} >{\centering\arraybackslash}m{0.8cm} >{\centering\arraybackslash}m{0.9cm} >{\centering\arraybackslash}m{0.8cm} >{\centering\arraybackslash}m{0.8cm}
}
\toprule
Networks & ER & WS & BA & DG & GGDP & GW & Our \\
\midrule
LesMis\'{e}rables & 0.2817 & 0.3002 & 0.3349 & 0.1226 & 0.6728 & 0.2396 & 0.0993 \\
RealityMining & 0.4215 & 0.4246 & 0.3969 & 0.2545 & 0.4324 & 0.5811 & 0.2298 \\
EmailMC & 0.4838 & 0.5131 & 0.3167 & 0.2971 & 0.5662 & 0.4255 & 0.1882 \\
Jazz & 0.2639 & 0.3692 & 0.2311 & 0.1076 & 0.6728 & 0.4246 & 0.0847 \\
Celegans & 0.1472 & 0.2324 & 0.1825 & 0.1300 & 0.6875 & 0.5198 & 0.0558 \\
EmailEC & 0.4232 & 0.4977 & 0.2893 & 0.1095 & 0.6931 & 0.3464 & 0.0791 \\
EmailRVU & 0.1969 & 0.3273 & 0.2236 & 0.0194 & 0.6931 & 0.4565 & 0.0145 \\
Blogs & 0.4254 & 0.5060 & 0.3142 & 0.0766 & 0.6931 & 0.3008 & 0.0560 \\
IFM & 0.2633 & 0.3717 & 0.2543 & 0.0277 & 0.6931 & 0.3830 & 0.0261 \\
AmazonMTurk & 0.3488 & 0.4641 & 0.3682 & 0.1008 & 0.6931 & 0.2310 & 0.0696 \\
CANetSci & 0.0492 & 0.1693 & 0.1155 & 0.0278 & 0.6931 & 0.4138 & 0.0087 \\
Bible & 0.2457 & 0.3663 & 0.2312 & 0.0538 & 0.6931 & - & 0.0570 \\
Yeast & 0.0739 & 0.0831 & 0.0136 & 0.0474 & - & - & 0.0048 \\
HumanProteins2 & 0.2588 & 0.3948 & 0.3045 & 0.0459 & 0.6931 & 0.2290 & 0.0324 \\
SocHamsterster & 0.3001 & 0.3999 & 0.2692 & 0.0437 & - & 0.4450 & 0.0206 \\
HumanProteins1 & 0.1310 & 0.2439 & 0.1658 & 0.0201 & - & - & 0.0072 \\
HsLc & 0.3700 & 0.4606 & 0.3035 & - & - & 0.3492 & 0.0315 \\
CAGrQc & 0.1674 & 0.3358 & 0.2100 & - & - & - & 0.0107 \\
MovieLens & 0.5459 & 0.5852 & 0.2907 & 0.0990 & - & 0.4996 & 0.0828 \\
Advogato & 0.3757 & 0.4713 & 0.3366 & - & - & 0.5103 & 0.0419 \\
Dmela & 0.2472 & 0.4084 & 0.2363 & - & - & - & 0.0136 \\
LastFmAsia & 0.2118 & 0.3612 & 0.2462 & - & - & - & 0.0074 \\
Facebook & 0.3113 & 0.4231 & 0.2823 & - & - & - & 0.0105 \\
CAAstroPh1 & 0.3437 & 0.4457 & 0.2870 & - & - & - & 0.0129 \\
CAAstroPh2 & 0.3827 & 0.4730 & 0.2981 & - & - & - & 0.0154 \\
Marvel & 0.2102 & 0.3258 & 0.2163 & - & - & 0.4463 & 0.0076 \\
CitCora & 0.1643 & 0.3023 & 0.1882 & - & - & - & 0.0020 \\
CitHepPh2 & 0.3476 & 0.4450 & 0.2459 & - & - & - & 0.0045 \\
CAConMat & 0.1769 & 0.3121 & 0.1859 & - & - & - & 0.0038 \\
CitHepPh1 & 0.3161 & 0.4119 & 0.2151 & - & - & - & 0.0025 \\
Brightkite & 0.2964 & 0.4326 & 0.3186 & - & - & - & 0.0231 \\
Twitter & 0.4398 & 0.5118 & 0.2943 & - & - & - & 0.0101 \\
\bottomrule
\end{tabular}
\begin{tablenotes}
    \footnotesize
    \item[*] The JS metric denotes the Jensen-Shannon divergence.
    \item[*] The symbol `-' denotes cases where learning-based models are unable to perform computations even under the conditions of our high-performance computing platform, particularly for large networks.
\end{tablenotes}
\end{minipage}
} 
\end{table}

\clearpage
\section{Adaptability of Vari-linear Model to Properties of Classical Networks}\label{A-cn}

\setcounter{table}{0} 

\begin{table}[h]
\centering
\captionsetup{justification=raggedright,singlelinecheck=false} 
\caption{Comparison of properties between vari-linear network and classical networks.}
\label{tab-cn}
\begin{tabular}{ l >{\centering\arraybackslash}m{0.8cm} >{\centering\arraybackslash}m{0.8cm} >{\centering\arraybackslash}m{0.8cm} >{\centering\arraybackslash}m{1.2cm} >{\centering\arraybackslash}m{0.8cm} >{\centering\arraybackslash}m{0.8cm} >{\centering\arraybackslash}m{1.2cm} }
\toprule
Model & $N$ & $M$ & $\langle k \rangle$ & $\langle C \rangle$ & $\langle L \rangle$ & $D$ & $\rho$ \\
\midrule
ER\footnotemark[1] & 200 & 597 & 6.0 & 2.9$\times10^{-2}$ & 3.15 & 6.07 & 3.00$\times10^{-2}$ \\
Our$\#1$\footnotemark[2] & 200 & 600 & 6.0 & 4.5$\times10^{-2}$ & 3.28 & 6.88 & 3.02$\times10^{-2}$ \\
WS\footnotemark[3] & 200 & 600 & 6.0 & 3.2$\times10^{-1}$ & 3.76 & 6.76 & 3.02$\times10^{-2}$ \\
Our$\#2$\footnotemark[4] & 200 & 600 & 6.0 & 1.9$\times10^{-1}$ & 3.86 & 8.29 & 3.02$\times10^{-2}$ \\
BA\footnotemark[5] & 200 & 591 & 5.9 & 9.5$\times10^{-2}$ & 2.88 & 5.00 & 2.97$\times10^{-2}$ \\
Our$\#3$\footnotemark[6] & 200 & 600 & 6.0 & 1.1$\times10^{-1}$ & 2.96 & 6.12 & 3.02$\times10^{-2}$ \\
\bottomrule     
\end{tabular}
\begin{tablenotes}
    \footnotesize
    \item[*] All results are based on $10,000$ Monte Carlo simulations, $N$ denotes the number of nodes, $M$ denotes the number of edges, $\langle k \rangle$ denotes the average degree, $\langle C \rangle$ denotes the average clustering coefficient, $\langle L \rangle$ denotes the average path length, $D$ denotes the network diameter, and $\rho$ denotes network density.
    \item[1] Parameter settings: $n=1000$, $p=6\times10^{-3}$.
    \item[2] Parameter settings: $n=1000$, $k=6.126$, $r=-1.5$.
    \item[3] Parameter settings: $n=1000$, $k=6$, $p=0.2$.
    \item[4] Parameter settings: $n=1000$, $k=6.126$, $r=-10$.
    \item[5] Parameter settings: $n=1000$, $m=3$.
    \item[6] Parameter settings: $n=1000$, $k=6.126$, $r=1$.  
\end{tablenotes}
\end{table}

\clearpage
\section{Numerical Results of Ablation Experiment}\label{A-ae}

\setcounter{table}{0} 

\begin{table}[h]
\centering
\captionsetup{justification=raggedright,singlelinecheck=false} 
\rotatebox{90}{ 
\resizebox{0.8\textheight}{!}{ 
\begin{minipage}{\textheight}  
\centering
\captionof{table}{Numerical Results of Ablation Experiment}
\label{tab-ae}
\begin{tabular}{
l >{\centering\arraybackslash}m{0.8cm} >{\centering\arraybackslash}m{0.8cm} >{\centering\arraybackslash}m{0.8cm} >{\centering\arraybackslash}m{0.8cm} >{\centering\arraybackslash}m{0.8cm} >{\centering\arraybackslash}m{0.8cm} >{\centering\arraybackslash}m{0.8cm} >{\centering\arraybackslash}m{0.8cm} >{\centering\arraybackslash}m{0.8cm} >{\centering\arraybackslash}m{0.62cm} >{\centering\arraybackslash}m{0.62cm} >{\centering\arraybackslash}m{0.62cm} >{\centering\arraybackslash}m{0.8cm} >{\centering\arraybackslash}m{0.8cm} >{\centering\arraybackslash}m{0.8cm} >{\centering\arraybackslash}m{0.8cm} >{\centering\arraybackslash}m{0.8cm} >{\centering\arraybackslash}m{0.8cm}
}
\toprule
\multirow{3}{*}{Networks}  & \multicolumn{3}{c}{NLSD} & \multicolumn{3}{c}{SINS} & \multicolumn{3}{c}{GDD} & \multicolumn{3}{c}{WD} & \multicolumn{3}{c}{KS} & \multicolumn{3}{c}{JS} \\
\cmidrule(lr){2-4} \cmidrule(lr){5-7} \cmidrule(lr){8-10} \cmidrule(lr){11-13} \cmidrule(lr){14-16} \cmidrule(lr){17-19} 
 & EPG & NPA & Our & EPG & NPA & Our & EPG & NPA & Our & EPG & NPA & Our & EPG & NPA & Our & EPG & NPA & Our \\
\midrule
LesMis\'{e}rables & 0.1114 & 0.0892 & 0.0831 & 20.513 & 14.135 & 12.596 & 0.5714 & 0.7403 & 0.7273 & 4.00 & 2.23 & 0.65 & 0.4286 & 0.2597 & 0.0909 & 0.1709 & 0.1930 & 0.0993 \\
RealityMining & 0.1483 & 0.0111 & 0.0012 & 24.627 & 15.510 & 7.532 & 0.4743 & 0.7760 & 0.7292 & 43.43 & 18.88 & 12.08 & 0.8851 & 0.5000 & 0.2604 & 0.5929 & 0.3617 & 0.2298 \\
EmailMC & 0.0337 & 0.0089 & 0.0007 & 20.977 & 15.647 & 7.937 & 0.7036 & 0.7545 & 0.7784 & 28.84 & 16.79 & 6.10 & 0.5749 & 0.2934 & 0.1198 & 0.3553 & 0.3142 & 0.1882 \\
Jazz & 0.0190 & 0.0260 & 0.0042 & 21.157 & 14.443 & 7.417 & 0.6535 & 0.7980 & 0.8460 & 19.68 & 6.76 & 1.94 & 0.6157 & 0.1818 & 0.0606 & 0.2804 & 0.1684 & 0.0847 \\
Celegans & 0.0345 & 0.0042 & 0.0010 & 18.908 & 10.134 & 8.607 & 0.6094 & 0.7694 & 0.7290 & 9.78 & 2.53 & 2.22 & 0.6061 & 0.1717 & 0.0774 & 0.2628 & 0.1001 & 0.0558 \\
EmailEC & 0.0889 & 0.0867 & 0.0792 & 18.025 & 13.953 & 10.186 & 0.8796 & 0.9308 & 0.9303 & 37.04 & 21.67 & 4.63 & 0.4318 & 0.3005 & 0.0786 & 0.1665 & 0.2608 & 0.0791 \\
EmailRVU & 0.0073 & 0.0041 & 0.0008 & 20.075 & 13.473 & 10.719 & 0.7321 & 0.8839 & 0.8892 & 5.96 & 2.54 & 0.63 & 0.4078 & 0.2745 & 0.0485 & 0.1260 & 0.1811 & 0.0145 \\
Blogs & 0.0131 & 0.0146 & 0.0058 & 22.442 & 13.697 & 9.890 & 0.8986 & 0.8950 & 0.9109 & 19.19 & 12.99 & 4.01 & 0.3405 & 0.2761 & 0.0915 & 0.1088 & 0.1724 & 0.0560 \\
IFM & 0.0096 & 0.0016 & 0.0005 & 17.047 & 10.269 & 3.663 & 0.9096 & 0.9313 & 0.9491 & 6.54 & 3.71 & 1.00 & 0.3444 & 0.2378 & 0.0490 & 0.0944 & 0.1191 & 0.0261 \\
AmazonMTurk & 0.0319 & 0.0567 & 0.0513 & 20.151 & 16.044 & 13.067 & 0.9168 & 0.9496 & 0.9352 & 4.46 & 3.92 & 2.09 & 0.1980 & 0.2253 & 0.1836 & 0.0571 & 0.0888 & 0.0696 \\
CANetSci & 0.0276 & 0.0483 & 0.0362 & 19.891 & 17.201 & 16.124 & 0.6828 & 0.8850 & 0.8908 & 1.69 & 0.37 & 0.20 & 0.4162 & 0.1752 & 0.0287 & 0.1145 & 0.0720 & 0.0087 \\
Bible & 0.0248 & 0.0319 & 0.0411 & 20.877 & 14.885 & 11.535 & 0.9357 & 0.9738 & 0.9780 & 6.71 & 3.08 & 0.91 & 0.3801 & 0.2059 & 0.1049 & 0.1878 & 0.0822 & 0.0570 \\
Yeast & 0.0046 & 0.0003 & 0.0089 & 16.433 & 16.816 & 11.438 & 0.8658 & 0.8468 & 0.8481 & 0.54 & 0.44 & 0.08 & 0.1139 & 0.0401 & 0.0160 & 0.0136 & 0.0064 & 0.0048 \\
HumanProteins2 & 0.0232 & 0.0311 & 0.0278 & 19.749 & 15.506 & 8.722 & 0.9533 & 0.9837 & 0.9866 & 3.13 & 2.70 & 1.00 & 0.1599 & 0.1644 & 0.1050 & 0.0345 & 0.0625 & 0.0324 \\
SocHamsterster & 0.0253 & 0.0345 & 0.0310 & 21.024 & 17.763 & 13.800 & 0.8753 & 0.9446 & 0.9518 & 9.13 & 5.11 & 1.28 & 0.3504 & 0.2465 & 0.0334 & 0.0986 & 0.1308 & 0.0206 \\
HumanProteins1 & 0.0120 & 0.0220 & 0.0178 & 16.290 & 14.453 & 16.239 & 0.9263 & 0.9133 & 0.8886 & 2.08 & 1.09 & 0.25 & 0.2777 & 0.3329 & 0.0243 & 0.0578 & 0.0739 & 0.0072 \\
HsLc & 0.0115 & 0.0129 & 0.0106 & 22.942 & 15.658 & 15.302 & 0.9345 & 0.9681 & 0.8608 & 13.05 & 9.56 & 3.99 & 0.3004 & 0.2560 & 0.0764 & 0.0816 & 0.1523 & 0.0315 \\
CAGrQc & 0.0103 & 0.0150 & 0.0132 & 18.894 & 17.319 & 20.935 & 0.8439 & 0.8123 & 0.8136 & 2.93 & 1.69 & 0.69 & 0.2730 & 0.3392 & 0.1578 & 0.0624 & 0.0535 & 0.0107 \\
MovieLens & 0.0002 & 0.0001 & 0.0003 & 22.567 & 15.988 & 7.481 & 0.9184 & 0.9503 & 0.9600 & 246.97 & 125.41 & 17.49 & 0.4551 & 0.2717 & 0.0452 & 0.2212 & 0.2710 & 0.0828 \\
Advogato & 0.0109 & 0.0127 & 0.0112 & 20.422 & 14.171 & 17.552 & 0.9816 & 0.9484 & 0.9493 & 8.64 & 6.92 & 3.07 & 0.2159 & 0.2077 & 0.1055 & 0.0688 & 0.0891 & 0.0419 \\
Dmela & 0.0019 & 0.0023 & 0.0012 & 15.638 & 10.128 & 5.230 & 0.9223 & 0.9756 & 0.8998 & 4.28 & 2.81 & 1.29 & 0.2739 & 0.2226 & 0.0526 & 0.0669 & 0.0838 & 0.0136 \\
LastFmAsia & 0.0028 & 0.0057 & 0.0042 & 19.891 & 17.694 & 15.699 & 0.9368 & 0.9745 & 0.8970 & 4.18 & 2.41 & 0.70 & 0.2811 & 0.1886 & 0.0299 & 0.0629 & 0.0905 & 0.0074 \\
Facebook & 0.0033 & 0.0046 & 0.0037 & 21.755 & 17.113 & 16.320 & 0.9208 & 0.9018 & 0.9027 & 7.50 & 5.22 & 2.12 & 0.2744 & 0.2253 & 0.0420 & 0.0638 & 0.1014 & 0.0105 \\
CAAstroPh1 & 0.0042 & 0.0051 & 0.0043 & 22.288 & 19.049 & 20.170 & 0.8262 & 0.9281 & 0.9285 & 10.02 & 6.07 & 1.93 & 0.3182 & 0.5183 & 0.0513 & 0.0896 & 0.1026 & 0.0129 \\
CAAstroPh2 & 0.0042 & 0.0044 & 0.0040 & 22.332 & 18.605 & 19.276 & 0.8547 & 0.9421 & 0.9373 & 14.53 & 9.28 & 3.10 & 0.3240 & 0.2543 & 0.0580 & 0.0903 & 0.1466 & 0.0154 \\
Marvel & 0.0035 & 0.0032 & 0.0028 & 24.741 & 22.366 & 16.286 & 0.9839 & 0.9943 & 0.9977 & 6.36 & 3.29 & 1.18 & 0.3508 & 0.1935 & 0.0241 & 0.0812 & 0.2008 & 0.0076 \\
CitCora & 0.0012 & 0.0022 & 0.0018 & 21.126 & 18.542 & 17.057 & 0.9547 & 0.9356 & 0.9356 & 4.60 & 1.71 & 0.17 & 0.3901 & 0.2046 & 0.0101 & 0.1060 & 0.1720 & 0.0020 \\
CitHepPh2 & 0.0028 & 0.0030 & 0.0028 & 21.711 & 17.868 & 17.882 & 0.9760 & 0.9855 & 0.9863 & 17.91 & 8.45 & 0.98 & 0.4133 & 0.2526 & 0.0145 & 0.1229 & 0.2262 & 0.0045 \\
CAConMat & 0.0016 & 0.0024 & 0.0021 & 21.087 & 17.977 & 16.859 & 0.8958 & 0.8810 & 0.9664 & 4.81 & 1.91 & 0.29 & 0.3994 & 0.2201 & 0.0302 & 0.1159 & 0.1723 & 0.0038 \\
CitHepPh1 & 0.0022 & 0.0024 & 0.0022 & 22.555 & 17.419 & 15.661 & 0.9287 & 0.9067 & 0.9074 & 17.38 & 7.19 & 0.21 & 0.4760 & 0.3961 & 0.0041 & 0.1572 & 0.1941 & 0.0025 \\
Brightkite & 0.0009 & 0.0012 & 0.0012 & 18.731 & 15.989 & 15.398 & 0.9937 & 0.9877 & 0.9877 & 4.25 & 3.83 & 2.07 & 0.1653 & 0.1702 & 0.0990 & 0.0309 & 0.0634 & 0.0231 \\
Twitter & 0.0008 & 0.0008 & 0.0007 & 23.024 & 20.557 & 22.706 & 0.9906 & 0.9790 & 0.9793 & 31.46 & 20.42 & 5.59 & 0.3431 & 0.2613 & 0.0484 & 0.0927 & 0.2811 & 0.0101 \\
\bottomrule     
\end{tabular}
\begin{tablenotes}
    \footnotesize
    \item[*] For the GDD metric, higher values indicate better method performance. For other metrics, lower values indicate better method performance.
\end{tablenotes}
\end{minipage}
} 
} 
\end{table}

\end{appendices}
\clearpage


\end{document}